\documentclass[aps,pra,twocolumn,english]{revtex4-1}

\usepackage{graphicx}
\usepackage{amsmath}
\usepackage{amssymb}
\usepackage{dsfont}
\usepackage{fancyhdr}
\usepackage{yfonts}

\usepackage[utf8]{inputenc}

\newcommand{\HT}{\textnormal{\textsc{HT}}}
\newcommand{\LT}{\textnormal{\textsc{LT}}}

\newcommand{\ULT}{\textnormal{\textsc{ULT}}}

\newcommand{\bra}[1]{\langle #1|}
\newcommand{\ket}[1]{|#1\rangle}
\newcommand{\braket}[2]{\langle #1|#2\rangle}

\DeclareFontFamily{U}{BOONDOX-calo}{\skewchar\font=45 }
\DeclareFontShape{U}{BOONDOX-calo}{m}{n}{
  <-> s*[1.05] BOONDOX-r-calo}{}
\DeclareFontShape{U}{BOONDOX-calo}{b}{n}{
  <-> s*[1.05] BOONDOX-b-calo}{}
\DeclareMathAlphabet{\mathcalboondox}{U}{BOONDOX-calo}{m}{n}
\SetMathAlphabet{\mathcalboondox}{bold}{U}{BOONDOX-calo}{b}{n}
\DeclareMathAlphabet{\mathbcalboondox}{U}{BOONDOX-calo}{b}{n}

\usepackage{hyperref}
\hypersetup{
  colorlinks   = true, 
  urlcolor     = blue, 
  linkcolor    = blue,
  citecolor   = blue,
}

\begin{document}
\title{High Finesse Fiber Fabry-Perot Cavities: \\ Stabilization and Mode Matching Analysis}
\author{J. Gallego, S. Ghosh, S. K. Alavi, W. Alt, M. Martinez-Dorantes, D. Meschede, L. Ratschbacher}\email[]{lratschb@uni-bonn.de}
\affiliation{Institut f\"ur Angewandte Physik der Universit\"at Bonn, Wegelerstrasse 8, 53115 Bonn, Germany}

\begin{abstract}  
Fiber Fabry-Perot cavities, formed by micro-machined mirrors on the end-facets of optical fibers, are used in an increasing number of technical and scientific applications, where they typically require precise stabilization of their optical resonances. Here, we study two different approaches to construct fiber Fabry-Perot resonators and stabilize their length for experiments in cavity quantum electrodynamics with neutral atoms. A piezo-mechanically actuated cavity with feedback based on the Pound-Drever-Hall locking technique is compared to a novel rigid cavity design that makes use of the high passive stability of a monolithic cavity spacer and employs thermal self-locking and external temperature tuning. Furthermore, we present a general analysis of the mode matching problem in fiber Fabry-Perot cavities, which explains the asymmetry in their reflective line shapes and has important implications for the optimal alignment of the fiber resonators. Finally, we  discuss the issue of fiber-generated background photons.  We expect that our results contribute towards the integration of high-finesse fiber Fabry-Perot cavities into compact and robust quantum-enabled devices in the future. 
\end{abstract}
\maketitle
\section{Introduction}
In recent years optical high-finesse resonators with small mode volumes have become powerful tools for enhancing the interaction between light and matter. Resonator-enhanced interaction in Cavity Quantum Electrodynamics (CQED), for instance, provides the basis for the realization of efficient single photon interfaces in quantum communication and information~\cite{Kuhn2002,Ritter2012} and the study of quantum opto-mechanical systems~\cite{Purdy2010}. 

Among the geometries of optical micro-cavities that are currently investigated, optical Fiber Fabry-Perot Cavities (FFPCs)~\cite{Colombe2007} are particularly attractive for CQED experiments because they combine several desirable features.  Formed by dielectric mirrors on the end-facets of opposing optical glass fibers, FFPCs provide small mode volumes, high optical Q factors, direct access to the cavity mode and intrinsic fiber coupling of the mode field. Details of the fabrication~\cite{Muller2010, Hunger2012} and optical characterization of fiber mirrors and cavities~\cite{Hunger2010, Brandstatter2013}, including the effects of thermo-optical bistability~\cite{Hunger2010} and cavity polarization mode splitting~\cite{Takahashi2014, Uphoff2015}, have been described in several recent studies. To date fiber Fabry-Perot cavities have been successfully applied in experiments interfacing single photons with a wide range of quantum systems, including cold atoms~\cite{Colombe2007}, ions~\cite{Steiner2013}, and solid state emitters~\cite{Toninelli2010, Muller2009, Albrecht2013} as well as quantum opto-mechanical experiments~\cite{Flowers2012}. 

The resonator-enhanced light-matter interaction in CQED experiments relies on the precise tuning of a cavity resonance to an optical transition of the quantum system under investigation. In ever more miniaturized and integrated experimental setups the task of stabilizing high-finesse FFPCs to within a small fraction of their optical linewidth (corresponding to mirror displacements of order $10\,$pm) can pose significant challenges. In particular for future applications of high finesse micro-resonators, e.g. in portable devices that can operate outside of dedicated laboratory environments, novel technical solutions for assembly and stabilization will be required that are robust and compact.  

Additional challenges, which need to be considered for FFPCs but do not occur for cavities with macroscopic mirror substrates, arise from the direct fiber coupling of light to the resonator mode. For instance, due to the intrinsic fiber coupling, the efficiencies to couple light to and from the cavity mode are set by the relative alignment of the fiber mirrors. In order to achieve optimal cavity alignment it is therefore important to find strategies to accurately characterize the coupling efficiencies already during the assembling of the resonator.   

In this article we address the issue of cavity stabilization by studying two different architectures for high finesse fiber Fabry-Perot resonators. We compare a conventional, piezo-mechanically actuated FFPC with a novel rigid Fabry-Perot design. Furthermore we analyze the connection between the cavity alignment, cavity mode matching and reflective lineshapes for FFPCs and show how it can be used to achieve optimal FFPC alignment. In addition we discuss the issue of fiber generated background counts, which can obscure weak single-photon signals in CQED applications. 

The paper is organized as follows. Section~\ref{sec:piezocavity} describes the design of the piezo-mechanically actuated resonator and reports on its performance in a cavity stabilization scheme for neutral atom CQED experiments. In Section~\ref{sec:FFPCeffects} we discuss several of the challenges specific to FFPCs. The results of the FFPC mode matching analysis and the investigation of the issue of fiber generated background counts are presented. The topic of Section~\ref{sec:rigidcavity} is the experimental realization of a rigid high finesse FFPC and the characterization of its thermal response for the purpose of cavity stabilization and tuning. 
In Section~\ref{sec:theorymodematching} the theory of reflective cavity mode matching in FFPCs is developed in detail, and its connection to FFPC alignment is shown both theoretically and experimentally.  
Section~\ref{sec:conclusion} concludes the paper with a comparison of the respective advantages of piezo-mechanically actuated and rigid FFPCs. 

%%%%%%%%%%%%%%%%%%%%%%%%%%%%%%%%%%%%%%%%%%%%%%%%%%%%%%%%%%%%%%%%
\begin{figure}[ht]
    \includegraphics[width=1.00\columnwidth]{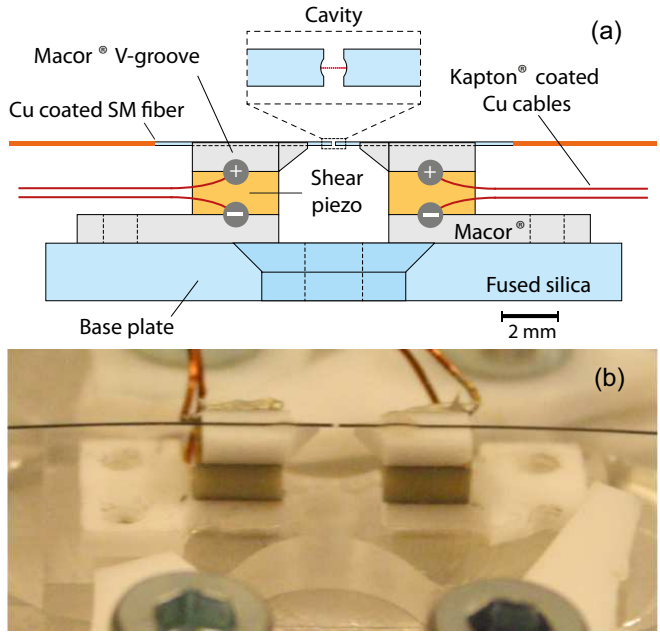}
  	\caption{Piezo-mechanically actuated fiber Fabry-Perot cavity assembly. {\bf a} Vertical section of the ultra-high-vacuum compatible cavity design. {\bf b} Image of the completed cavity assembly.}{\label{fig:piezocavity}}
\end{figure}
%%%%%%%%%%%%%%%%%%%%%%%%%%%%%%%%%%%%%%%%%%%%%%%%%%%%%%%%%%%%%%%%
\section{Piezo-mechanically actuated FFPC} \label{section:piezocavity}
\label{sec:piezocavity}
\subsection{Construction} 
\label{piezoconstruction}
\subsubsection*{Fiber mirror production}
The fiber mirrors used for all our cavities are fabricated by locally ablating material from the cleaved end-facets of copper-coated optical single mode fibers~\footnote{IVG\textsuperscript{\textregistered} Cu800.} using focused CO$_2$ laser light~\cite{Muller2010, Hunger2012}. We typically apply 50 laser pulses of $6\,$ms duration with a peak power of $20\,$W and $135\,\mu$m waist radius to generate Gaussian-shaped concave structures with low ellipticity and typical radii of curvature of $\sim 170\,\mu m$. After the application of a high-reflectivity dielectric coating (ion-beam sputtered multilayer SiO$_2$-Ta$_2$O$_5$), the fiber mirrors are thermally cured in air for $5\,$h at $300^{\circ}$C to reduce absorption losses in the coatings~\cite{Brandstatter2013}. For two different coating runs with high (HT) and low transmission (LT), the experimentally determined transmissions $\mathcal{T}_{\rm{HT}, 780\,\rm{nm}}=126\pm13\,$ppm, $\mathcal{T}_{\rm{HT}, 850\,\rm{nm}}=490 \pm 30\,$ppm, $\mathcal{T}_{\rm{LT}, 780\,\rm{nm}}=15\pm3\,$ppm, $\mathcal{T}_{\rm{LT}, 850\,\rm{nm}}=140 \pm 20\,$ppm and mirror losses $\mathcal{L}_{\rm{HT}, 780\,\rm{nm/} 850\,\rm{nm}} = 26\pm5\,$ppm, $\mathcal{L}_{\rm{LT}, 780\,\rm{nm/} 850\,\rm{nm}} = 25\pm5\,$ppm agree well with the theoretically predicted values at $780\,$nm and $850\,$nm wavelength, respectively.
\subsubsection*{Assembling the cavity}
The design and the completed assembly of the piezo-mechanically actuated fiber Fabry-Perot cavity are shown in Figure~\ref{fig:piezocavity}. The cavity is aligned by setting the two opposing fiber mirrors at the desired length of the cavity, rotating one fiber mirror around its axis to minimize the cavity polarization mode splitting~\cite{Takahashi2014, Uphoff2015} and varying the transverse mirror positions and angles to maximize the cavity to fiber mode coupling efficiency (see Section~\ref{sec:theorymodematching}). The fibers are then glued onto individual V-grooves including high-voltage shear piezos. The piezos allow the cavity length to be scanned over more than one free spectral range once the stacks have been glued onto a common base plate. To avoid misalignment of the cavity in the transversal directions during curing and vacuum bake-out we utilize UV curable glue~\footnote{EPO-TEK\textsuperscript{\textregistered} OG116-31} for the final bonding of the cavity and keep the thickness of glue layers to a minimum. 
\subsubsection*{Cavity parameters}
The piezo-mechanically actuated FFPC is built by combining a high and low transmission fiber mirror. The one-sided resonator maximizes the out-coupling efficiency of photons into the single spatial fiber mode at the high transmission side of the cavity. The parameters of the cavity at $780\,$nm wavelength are: cavity length $L_{\rm{cav}} = 93.3\,\mu$m, finesse $\mathcal{F}=32\,800\pm 1100$, cavity field decay rate $\kappa = 2\pi \cdot (24.5\pm 0.8)\,$MHz and polarization dependent mode splitting $\Delta\nu_{\rm{PMS}}=9.0\pm 0.3\,$MHz. From the observed dip in the reflected power at the cavity resonance we estimate the spatial mode matching between the cavity and the fiber of the HT mirror to be $\epsilon_{\rm{HT}} = 0.60\pm 0.02$ (see Sections~\ref{sec:FFPCeffects} and \ref{sec:theorymodematching}). 
%%%%%%%%%%%%%%%%%%%%%%%%%%%%%%%%%%%%%%%%%%%%%%%%%%%%%%%%%%%%%%%
\begin{figure*}
    \includegraphics[width=1.00\textwidth]{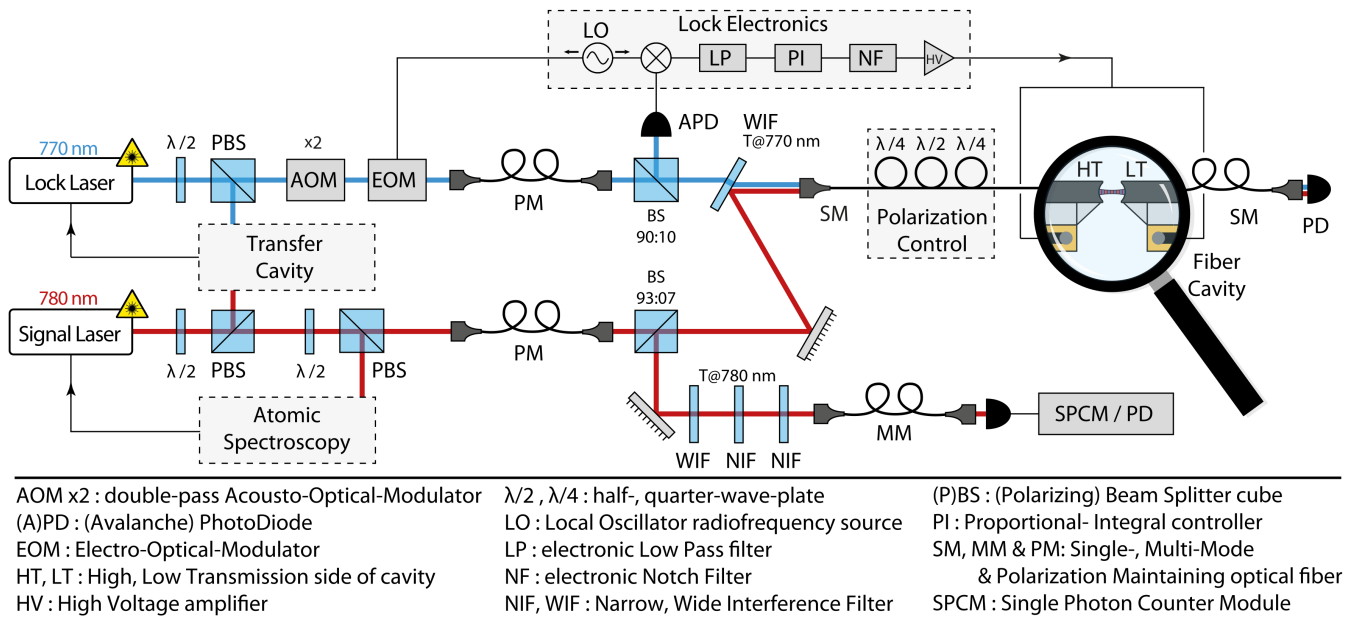}
  	\caption{Fiber cavity locking scheme for cavity QED experiments with neutral atoms. An optical resonance of the fiber resonator is stabilized at the rubidium D2-line at $780\,$nm by locking a cavity mode with a frequency separation of three free spectral ranges to a far detuned ``lock'' laser at $770\,$nm. The ``lock'' laser itself is stabilized via a transfer cavity to a rubidium-referenced ``signal'' laser that is used to resonantly address the cavity.  ``Signal'' and ``lock'' photons enter and exit the one-sided cavity through the fiber mirror with high transmission.  A series of band-pass interference filters is employed to separate the 1$\,\mu$W ``lock'' light from single photon signals at $780\,$nm prior to detection.} {\label{fig:lockingScheme}}
\end{figure*}
%%%%%%%%%%%%%%%%%%%%%%%%%%%%%%%%%%%%%%%%%%%%%%%%%%%%%%%%%%%%%%%
\subsection{Active stabilization} 
\begin{figure}
    \includegraphics[width=1.00\columnwidth]{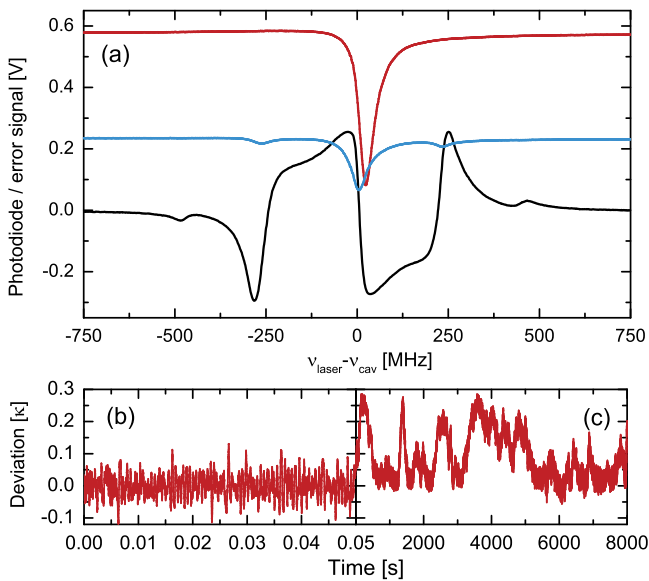}
	\caption[Level Scheme]{Stability of the piezo-mechanically actuated cavity. {\bf a} The Pound-Drever-Hall error signal (black, bottom) and the reflected ``lock''  (blue, middle) and ``signal'' power (red, top) are shown for a single scan of the cavity length. The powers incident on the cavity are $1.5\,\mu$W and a few $\mu$W for the ``lock'' and ``signal'' light, respectively. {\bf b} The short-term and {\bf c} long-term stability of the locked cavity are determined from an out-of-loop measurement (see text for details).}{\label{fig:PDHCharacterization}}
\end{figure}

Our method for locking the piezo-mechanically actuated FFPC, shown in Fig.~\ref{fig:lockingScheme}, closely follows previous schemes developed for atomic CQED setups with macroscopic mirror substrates~\cite{Mabuchi1999}. To enable efficient interaction between single resonant ``signal'' photons and rubidium atoms coupled to the cavity mode, the FFPC is stabilized to be resonant with the Rb D2 transition by means of the locking chain outlined in Fig.~\ref{fig:lockingScheme}. 
First, the ``signal'' external-cavity diode laser at 780$\,$nm wavelength is referenced onto a Doppler-free polarization spectroscopy in a rubidium vapor cell. The stability of this laser is transfered onto a ``lock'' laser near $770\,$nm by means of an actively stabilized Fabry-Perot transfer cavity with $1.5\,$m length and finesse 50. The frequency separation of the ``lock'' and the ``signal'' laser $\nu_{\rm{lock}} - \nu_{\rm{signal}}$ can be fine-tuned with an acousto-optical modulator (AOM) in double-pass configuration to exactly three free spectral ranges of the FFPC, such that both waves are simultaneously resonant with the cavity. The ``lock'' and the ``signal'' light fields are combined by an interference filter and enter the FFPC through the high-transmission fiber mirror. Sidebands at $248.2\,$MHz are electro-optically-modulated onto the ``lock'' light, which is detected in reflection by a resonant avalanche photodiode (APD) to generate an error signal with the Pound-Drever-Hall method (see Fig.~\ref{fig:PDHCharacterization}a). To improve the bandwidth of the lock the analog proportional-integral controller that supplies feedback to the shear piezos features passive electronic notch filters in order to compensate for mechanical resonances of the cavity assembly at 18 kHz and 70 kHz. The final unity-gain bandwidth of the lock is $\sim3\,$kHz.

The performance of the cavity stabilization system is summarized in Fig.~\ref{fig:PDHCharacterization}b. For the purpose of the measurement, the cavity was resonantly stabilized to the frequency of the ``lock'' laser. The frequency of the ``signal'' laser was adjusted to the slope of the optical resonance, such that its reflected power yields a steep monotonic signal proportional to frequency deviations of the cavity in an out-of-loop measurement. The stabilized FFPC features frequency noise of less than $0.04 \, \kappa$ ($1\,$MHz) RMS on short time scales (20$\,$Hz to 2$\,$MHz bandwidth) and peak-to-peak drifts of less than $0.36\, \kappa$ ($8.8\,$MHz) over the course of $2\,$h. The dominant source of drifts of the locked cavity is the fluctuation of the FFPC relative to the ``lock'' laser, probably caused by residual interferences in the fiber (see next Section). The presented stability measurements have been performed in air on top of a standard optical table. We expect further improvements in the lock performance for FFPCs operated in vacuum and mounted on dedicated vibration isolation supports.      
%%%%%%%%%%%%%%%%%%%%%%%%%%%%%%%%%%%%%%%%%%%%%%%%%%%%%%%%%%%%%%%
%%%%%%%%%%%%%%%%%%%%%%%%%%%%%%%%%%%%%%%%%%%%%%%%%%%%%%%%%%%%%%%
\section{Effects specific to FFPCs}
\label{sec:FFPCeffects}
The direct coupling of light from optical fibers to the cavity mode and vice versa is an important feature of FFPCs that can strongly benefit integration and miniaturization of optical cavities. The intrinsic fiber connection, however, can also give rise to adverse effects that do not occur for macroscopic Fabry-Perot cavities coupled by free-space beams.   
\subsubsection*{Optical interference due to fiber coupler}
A first, simple example for the experimental challenges specific to FFPCs is the avoidance of reflections when coupling ``lock'' light into the FFPC's optical fiber. Optical interferences along the beam path between the electro-optical modulator, the resonator and the photo detector can lead to time-varying offsets in the Pound-Drever-Hall error signal~\cite{Whittaker1985} and thereby cause drifts in the cavity lock. We have tested several different methods to minimize optical reflections at the fiber in-coupler, including angle cleaving and manual FC/APC connectorization. The best results have been obtained by fusion splicing the FFPC single mode fibers to commercially connectorized FC/APC single mode fibers. 
\subsubsection*{Mode matching and asymmetric line shapes of FFPCs}
An important difference between FFPCs and traditional cavities concerns the coupling of light to the resonator mode. For FFPCs the spatial mode matching between the cavity mode and the incoming and outgoing fiber guided light fields are intrinsically determined by the relative alignment of the fiber mirrors. Once the fiber mirrors have been fixed in position, the coupling into the cavity cannot be altered any more. The two spatial mode matching efficiencies $\epsilon_{\rm{HT}}$ and $\epsilon_{\rm{LT}}$ are important figures of merit that enter the resonant transmission efficiency of a FFPC
\begin{equation}
\frac{P_{\rm{T,max}}}{P_{\rm{in}}} = \epsilon_{\rm{HT}} \,\epsilon_{\rm{LT}} \cdot \frac{4\, \mathcal{T}_{\rm{LT}}\mathcal{T}_{\rm{HT}}}{(\mathcal{T}_{\rm{HT}}+\mathcal{T}_{\rm{LT}}+\mathcal{L}_{\rm{HT}}+\mathcal{L}_{\rm{LT}})^2}
\label{eq:transmission}
\end{equation}
and determine the probability to retrieve a cavity photon 
\begin{equation}
p_{i,\rm{out}} = \epsilon_{i} \cdot \frac{ \mathcal{T}_{i}}{\mathcal{T}_{\rm{HT}}+\mathcal{T}_{\rm{LT}}+\mathcal{L}_{\rm{HT}}+\mathcal{L}_{\rm{LT}}}
\label{eq:collection}
\end{equation}
at the fiber output $i =\,$ HT or LT.
For cavity-based light-matter interfaces, $\epsilon_{i}$ furthermore sets an upper limit to the efficiency of mapping photons from mirror side $i$ into the cavity-coupled quantum system.
Many applications of FFPCs therefore rely on a method to determine and optimize a particular mode matching efficiency (in our case $\epsilon_{\rm{HT}}$) during the alignment of the cavity. 

Measurements of the FFPC transmission depend according to Eq.~\ref{eq:transmission} on the product $\epsilon_{\rm{HT}} \cdot \epsilon_{\rm{LT}}$ and can therefore not be applied to separately determine the mode matching efficiencies of FFPCs consisting of two single mode fibers. Naturally the simple transmission method also fails for one-sided FFPCs with one non-transparent mirror substrate.

Reflection measurements, on the other hand, offer the possibility to characterize and optimize the individual mode matching efficiencies. Their evaluation, however, requires a more careful analysis than has previously been considered, in order to extract $\epsilon_{i}$ from the observed drop in the power of the reflected light on resonance. 
The treatment of the reflective coupling problem in terms of the interference of spatially mode filtered light fields (see Section~\ref{sec:theorymodematching} and Supplemental Material~\footnote{The Supplemental Material is included in the source code of the article that can be downloaded from its arXiv page.} for details and~\cite{Wieman1982,Shaddock1999} for earlier related works on the topic) gives rise to two important practical results for FFPCs: 
%%%%%%%%%%%%%%%%%%%%%%%%%%%%%%%%%%%%%%%%%%%%%%%%%%%%%%%%%%%%%%%%
\begin{figure}
    \includegraphics[width=1\columnwidth]{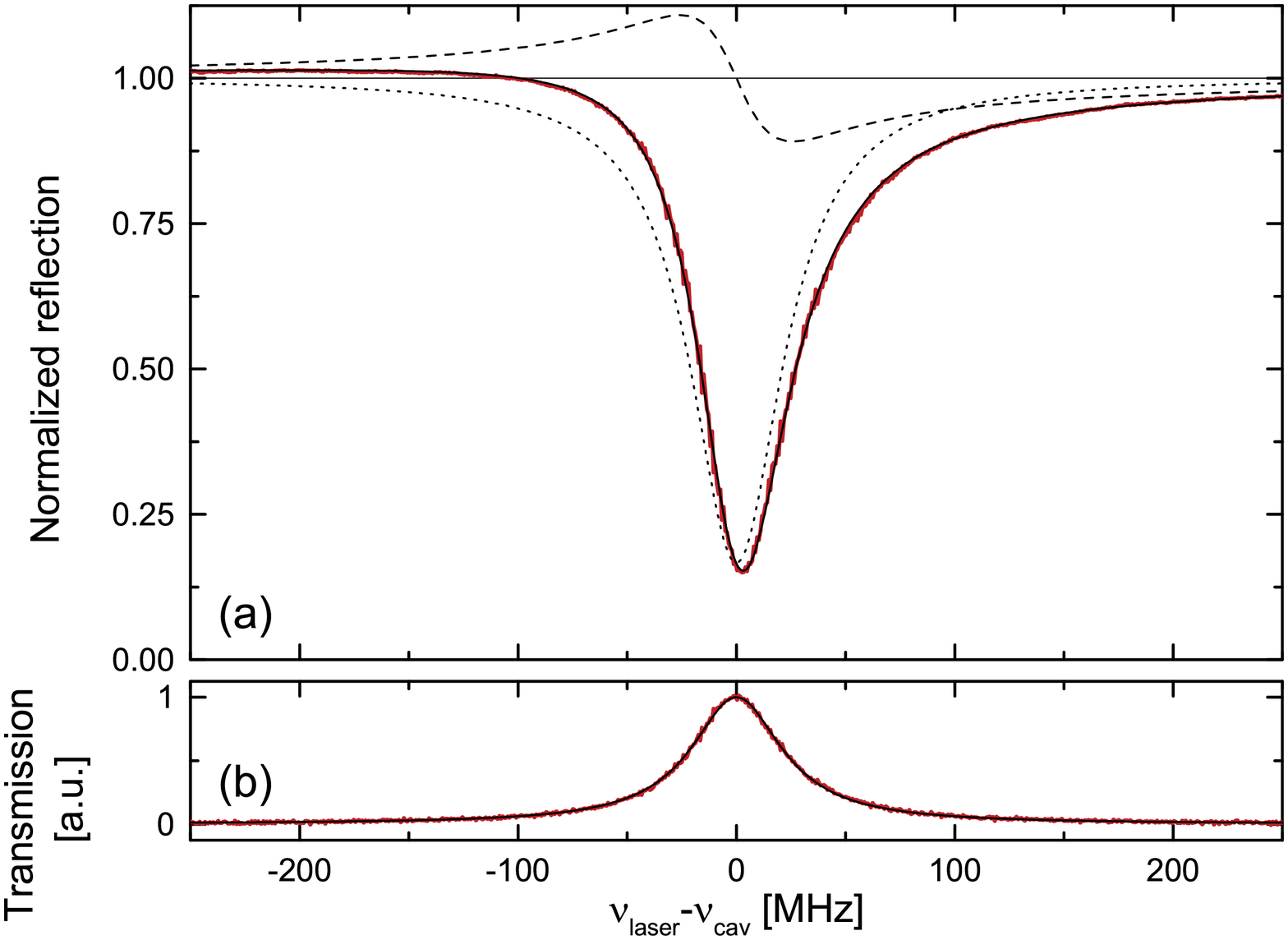}
  	\caption{Normalized reflective and transmittive line shapes of the FFPC (experimental data and fits).  {\bf a} The power and polarization independent asymmetry in the reflected light results from spatial mode filtering of the interfering fields reflected from the cavity by the single mode fiber (see Section~\ref{sec:theorymodematching}). The fit function (solid line, Eq.~\protect\ref{eq:asymmetricline} normalized to the off-resonantly detected power) predicted by the mode matching model is the sum of a Lorentzian (dotted line) and its corresponding dispersive function (dashed line). {\bf b} The transmitted power follows a Lorentzian curve.}{\label{fig:lineshape}}
\end{figure}
%%%%%%%%%%%%%%%%%%%%%%%%%%%%%%%%%%%%%%%%%%%%%%%%%%%%%%%%%%%%%%%%

First, our calculations show that for strongly overcoupled FFPCs, where the transmission of the incoupling mirror strongly exceeds the sum of all other contributions to the round trip losses ($\mathcal{T}_{\rm{HT}}\gg\mathcal{T}_{\rm{LT}}+\mathcal{L}_{\rm{LT}}+\mathcal{L}_{\rm{HT}}$), the alignment that maximizes the depth of reflected incoupling dip does not maximize the spatial mode matching efficiency $\epsilon_{\rm{HT}}$. In fact, for this experimentally very relevant type of one-sided cavity, the maximum depth of the incoupling dip will approach unity for misalignments that significantly reduce the mode matching efficiency. We experimentally confirm this result with measurements on a strongly overcoupled test cavity and find good agreement with the theoretical predictions (see Section~\ref{sec:theorymodematching}).    

Second, the model explains for the first time the origin of the asymmetry of FFPC resonance line shapes that are observed in reflection but not in transmission (see Fig.~\ref{fig:lineshape}). The asymmetry is found to be an intrinsic property of FFPCs and independent of the power and polarization of the incident light.
The asymmetric reflective line shapes are described by the sum of a Lorentzian and its corresponding dispersive function
\begin{align}
\label{eq:asymmetricline}
\frac{P_{\rm{out}}(v)}{P_{\rm{in}}}&=\eta_r -\eta_{\scriptscriptstyle {\mathcalboondox{L}}}\cdot \left(\frac{1}{1+v^2}- \mathcalboondox{A} \frac{v}{1+v^2}\right)\\
&{\rm with} \quad  v=2\pi(\nu_{\rm{laser}}-\nu_{\rm{cav}})/\kappa, \nonumber
\end{align}
where $\eta_{r}$ denotes the alignment-independent fraction of light that is detected in reflection for the off-resonant cavity.  Both the Lorentzian amplitude $\eta_{\scriptscriptstyle {\mathcalboondox{L}}}$ and the relative size of the dispersive component, which we introduce as the asymmetry parameter $\mathcalboondox{A}$, depend on the cavity geometry and alignment (see Section~\ref{sec:theorymodematching} for details). We find good agreement between theory and experimental measurements, thereby confirming our modeling of the effects of spatial mode filtering in FFPCs. 
\subsubsection*{Fiber generated background counts}
Finally, fiber generated background counts arise as an additional issue when employing FFPCs in CQED setups that operate at the single photon level. In these experiments single ``signal'' photons, which are resonant with the quantum system under investigation, need to be detected in the presence of the strong off-resonant ``lock'' light, which is needed for stabilizing the cavity (see Fig.~\ref{fig:lockingScheme}). In the case of cold neutral atoms the ``lock'' light furthermore acts as a optical dipole potential and determines the confinement of the neutral atoms. For the parameters of our piezo-actuated FFPC, an axial confinement of rubidium atoms of $1\,$mK depth corresponds to $1\,\mu$W of $770\,$nm ``lock'' light entering the incoupling fiber of the FFPC.
The separation of single ``signal'' photons from this strong ``lock'' light relies on spectral filtering, since the polarization degree of freedom is a resource needed for many CQED protocols. 

Experimentally, we find that after the suppression of ``lock'' laser photons at $770\,$nm by a series of band-pass interference filters (combined values: peak transmission $~90\,$\% and effective filter bandwidth of $\Delta \nu_{\rm{filter}} = 0.150\,$THz at $780\,$nm; extinction $>120\,$dB at $770\,$nm) the photon background is limited by frequency conversion inside the single mode fiber. Inelastic scattering of photons from $770\,$nm to $780\,$nm wavelength was measured to scale linearly in ``lock'' power $ P_{\rm{lock}}$ (in the range 1$\,\mu$W to 2$\,$mW) and fiber length $L_f$ (1-$10\,$m). The process happens at comparable rates both in the forward and backward direction, is spectrally uniform over the filter bandwidth around $780\,$nm and predominantly conserves the polarization. The rate of background photons for the $770\,$nm ``lock'' light is well described by $R_{\rm{bg}}=\sigma_{770\,\rm{nm},780\,\rm{nm}} \cdot P_{\rm{lock}} \cdot L_f \cdot \Delta \nu_{\rm{filter}}$, where $\sigma_{770\,\rm{nm},780\,\rm{nm}} = 0.004(2)\,$s$^{-1}$W$^{-1}$Hz$^{-1}$m$^{-1}$ is the spectral efficiency for shifting light at $770\,$nm to $780\,$nm wavelength. We have found similar background rates for red detuned ``lock'' light at $790\,$nm ($\sigma_{790\,\rm{nm},780\,\rm{nm}} = 0.002(1)\,$s$^{-1}$W$^{-1}$Hz$^{-1}$m$^{-1}$). For even further red detuned light at $850\,$nm, frequency conversion to $780\,$nm is energetically strongly suppressed ($\sigma_{850\,\rm{nm},780\,\rm{nm}} = 8(4)\cdot 10^{-7}\,$s$^{-1}$W$^{-1}$Hz$^{-1}$m$^{-1}$). However, in neutral atom CQED experiments the use of ``lock'' light with small frequency detuning is in many cases preferable, because it gives rise to similar coupling strengths to the ``signal'' mode for atoms trapped at different positions along the ``lock'' trap~\cite{Reimann2015}. 

The observed spectral conversion efficiencies for the three different pump wavelengths agree within a factor of two with values expected for spontaneous Raman scattering in optical fibers~\cite{Wardle1999}. We conclude that the background photon detection rate of $10^3\,$s$^{-1}$ in our setup is limited by Raman scattering and can therefore only be reduced by narrower spectral filtering and shorter optical fibers. 
\section{Rigid Fiber Fabry-Perot Cavities} \label{sec:rigidcavity}
Piezo-actuated designs, such as the cavity described in Section~\ref{sec:piezocavity}, have been the established solution for realizing high finesse FFPCs in recent experiments~\cite{Colombe2007, Steiner2013, Brandstatter2013}. The unique feature of this cavity design is the possibility to perform large, piezo-controlled scans of the cavity resonance, which can cover more than the free spectral range of the resonator. For many applications, however, FFPCs only require referencing to well-defined optical transition frequencies and could therefore benefit from alternative stabilization solutions that are potentially less complex, more robust, and better suited for miniaturization, integration and scalability. In the following we will present such a novel approach that is based on the high passive stability and precise thermal tunability of a rigid FFPC design. 
\subsection{Construction}
\subsubsection{Assembling the rigid cavity}
\begin{figure}    \includegraphics[width=1.00\columnwidth]{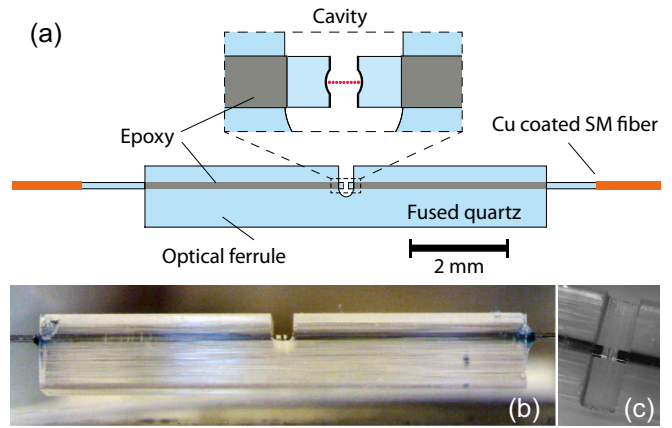}
	\caption{Rigid fiber Fabry-Perot cavity assembly. {\bf a} Vertical section of the cavity design. {\bf b} Side view of the suspended cavity assembly. {\bf c} Detailed view of the Fabry-Perot resonator.}{\label{fig:rigidcavitydesign}}
\end{figure}

The construction of our rigid FFPC, shown in Fig.~\ref{fig:rigidcavitydesign}, is inspired by the design of ultra-stable passive Fabry-Perot reference cavities~\cite{Martin2012}, where a constant cavity length is maintained by a monolithic cavity spacer with minimal thermal expansion and low vibrational sensitivity. To form the rigid cavity, two single mode mirror fibers with a cladding diameter of $125\,\mu m$ are inserted into a fused-quartz fiber ferrule with a bore of $131\,\mu$m in diameter. The ferrule ensures the transversal alignment of the two cavity fibers, while a $300\,\mu$m-wide cut at the center provides access to the cavity mode. The rotational degree of freedom can be used to maximize the fiber to cavity mode coupling for decentered fiber mirrors or alternatively to minimize the polarization mode splitting in the case of fibers with elliptical mirror shapes~\cite{Uphoff2015}. Precise longitudinal adjustment of the cavity length is performed by displacing one of the fibers with a piezo stage in order to set the cavity resonance to the desired frequency. The fibers are successively fixed in place by applying small amounts of glue to the optical fibers where they exit the ends of the ferrule. Due to capillary action, the low viscosity glue~\footnote{TRA-BOND\textsuperscript{\textregistered} F112} evenly fills the few $\mu$m wide gap between the ferrule bore and the inserted mirror fibers. In order to avoid misalignment during the room temperature curing of the glue, the cavity resonance is tracked by monitoring the transmission of a ``probe'' laser, and changes of cavity length are continuously compensated by the piezo stage until the glue becomes viscous. Typical frequency shifts during the hardening of the cavity that we have obtained with this method are $\sim 50\,$GHz. We believe that a further reduction of the misalignment could for example be achieved by filling the glue with SiO$_2$ nanospheres. Depending on the mounting positions of the cavity and the attached optical fibers, constant stress induced frequency shifts on the order of $10\,$GHz can be observed for the rigid cavity. To reduce the coupling to mechanical vibrations and external temperature fluctuations, we horizontally suspend the rigid cavity by its optical fibers (see Fig.~\ref{fig:rigidcavitydesign}).
\subsubsection*{Rigid cavity parameters}
Formed by two equal LT mirrors (see Section~\ref{sec:piezocavity}), the rigid FFPC has the following cavity parameters at $850\,$nm wavelength: cavity length $L_{\rm{cav}} = 47.96\pm0.01\,\mu$m, finesse $\mathcal{F}=19\,000\pm 2\,000$, cavity linewidth $\kappa = 2\pi \cdot(65\pm 6)\,$MHz and polarization dependent mode splitting $\Delta\nu_{\rm{PMS}}=38\pm 6\,$MHz. The transmission through the cavity and the observed depth of the cavity resonance in reflection are $P_{\rm{T,max}}/P_{\rm{in}} = 0.19\pm0.02$ and $\eta_{\rm{dip}} = 0.81\pm0.03$ respectively. 
%%%%%%%%%%%%%%%%%%%%%%%%%%%%%%%%%%%%%%%%%%%%%%%%%%%%%%%%%%%%%%%%    
%%%%%%%%%%%%%%%%%%%%%%%%%%%%%%%%%%%%%%%%%%%%%%%%%%%%%%%%%%%%%%%%
\subsection{Stability and tunability} %, 
\label{sec:stability}
The intrinsically rigid FFPC exhibits a high degree of passive stability. As a result of the monolithic design of the cavity spacer, which minimizes the sensitivity to mechanical vibrations, and the absence of noisy locking electronics, the short-term stability of the cavity length for the rigid FFPC (cf. Fig.~\ref{fig:closedloop}i) is comparable to the value obtained for the actively locked piezo-mechanically actuated FFPC.  

For the compensation of drifts and the fine-tuning of the resonance frequency, the rigid FFPC uniquely relies on the temperature response of the stiff cavity assembly. By placing the FFPC inside a temperature controlled enclosure, we find that homogeneous temperature changes shift the cavity resonance frequency by $\frac{d\nu_{\rm{cav}}}{dT}\sim 15\,$GHz/K. A theoretical estimate considering the linear expansion of the optical fibers, the cavity spacer and the coating predicts a value of $\frac{d\nu_{\rm{cav}}}{dT} \sim - 200 \,$MHz/K. We attribute the large observed temperature sensitivity to the influence of the connecting glue layer and expect that the thermal response to uniform temperature changes of the entire FFPC could be tuned by filling the glue in the future.
\subsubsection*{Active locking of the rigid FFPC with a heating laser}
An experimentally convenient way of controlling the temperature of a rigid cavity is by absorption of light from an external heating laser. The use of a heating laser enables strongly localized heating, requires only very limited amount of (optical) access and is compatible with ultra-high-vacuum setups. We characterize the technique by suspending the rigid FFPC on its fibers in vacuum and illuminating it with up to $20\,$mW of light at $405\,$nm wavelength. The laser light is focused onto the tip of one of the optical fibers forming the cavity, where it is efficiently absorbed by the thin black carbon layer present on the fiber (see Fig.~\ref{fig:rigidcavitydesign}). In this configuration the steady-state shift of the cavity resonance per incident laser power amounts to $\sim10\,$ GHz/mW, which allows the cavity to be tuned over more than 200 GHz. 
%%%%%%%%%%%%%%%%%%%%%%%%%%%%%%%%%%%%%%%%%%%%%%%%%%%%%%%%%%%%%%%%
\begin{figure}[ht]
    \includegraphics[width=1\columnwidth]{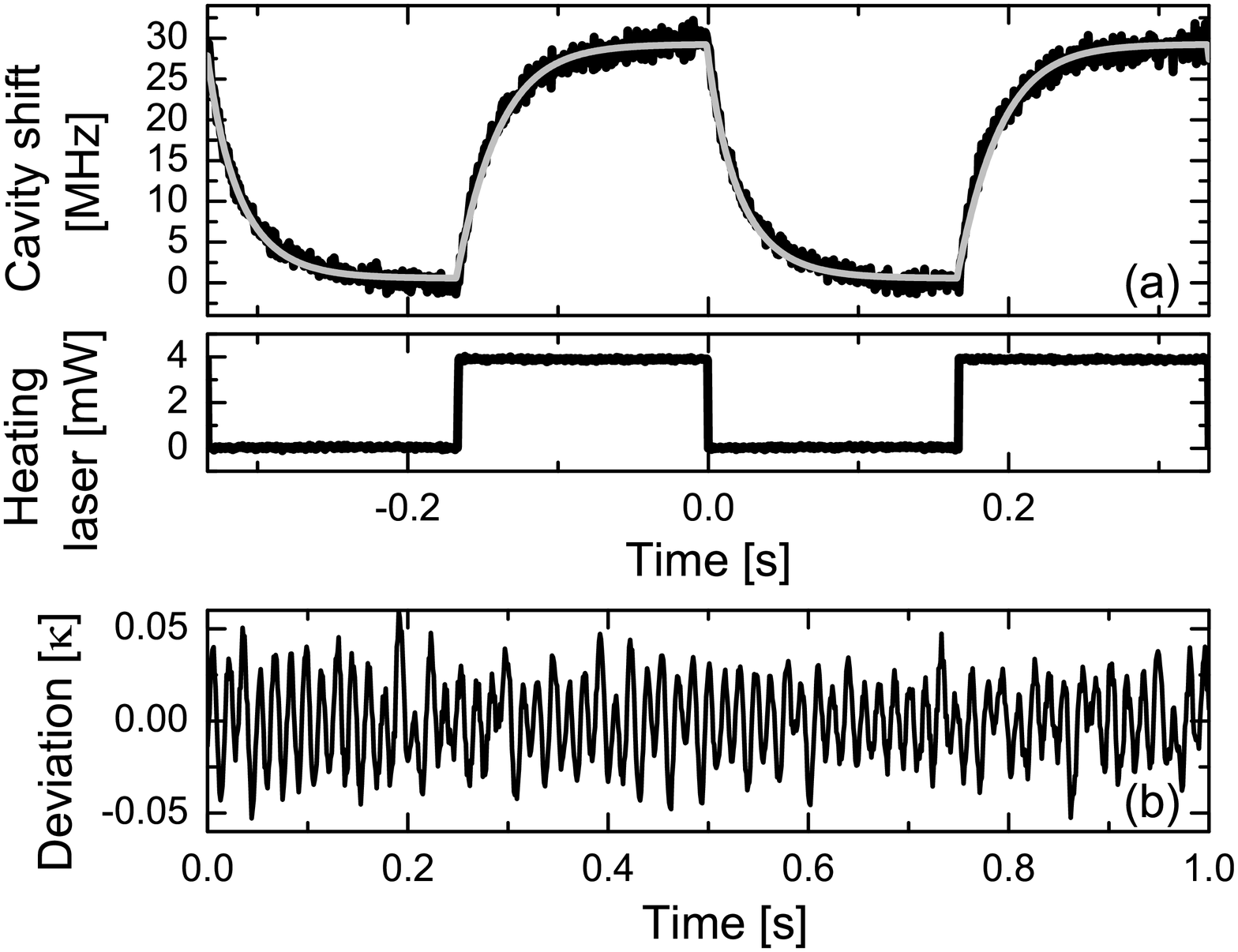} 
  	\caption{Thermal control of the FFPC with an external heating laser. {\bf a} Experimental open-loop response to a step change in external heating laser power (average over 32 repetitions). {\bf b} Stability of the cavity actively locked by an analog PI controller via external laser heating feedback (using the side of fringe transmission of a weak probe laser beam as the error signal). The discrete frequency component dominating the lock deviations results from mechanical noise at $\approx 65\,$Hz that is present on the optical table of the experiment.}{\label{fig:laserheating}} 
\end{figure}
%%%%%%%%%%%%%%%%%%%%%%%%%%%%%%%%%%%%%%%%%%%%%%%%%%%%%%%%%%%%%%%%

In order to go one step further and employ laser heating as an actuator for active control of the cavity resonance, we characterize the temporal response of the system. 
In Fig.~\ref{fig:laserheating}a we show the response of the cavity to a step change in the incident heating laser power. The fastest response with an exponential time constant of $\sim 25\,$ms and sensitivity on the $100\,$ms timescale of $\sim7$\,MHz/mW is achieved for defocused illumination (waist radius $\approx 1\,$mm) of the center of the ferrule. The cavity step response to external heating provides the information needed to design an optimized control system. As a proof of principle we realize a simple ``side-of-fringe'' lock of the cavity onto a stable ``probe'' laser. The transmission of the laser on the slope of the cavity resonance serves as the error signal input to an analog proportional-integral controller that provides feedback to the cavity via changes in the heating laser power. From the closed-loop error signal we estimate the RMS value of cavity lock deviations to be less than 0.025 $\kappa$ (see Fig.~\ref{fig:laserheating}b). 
%%%%%%%%%%%%%%%%%%%%%%%%%%%%%%%%%%%%%%%%%%%%%%%%%%%%%%%%%%%%%%%%
\begin{figure}[ht]
    \includegraphics[width=0.85\columnwidth]{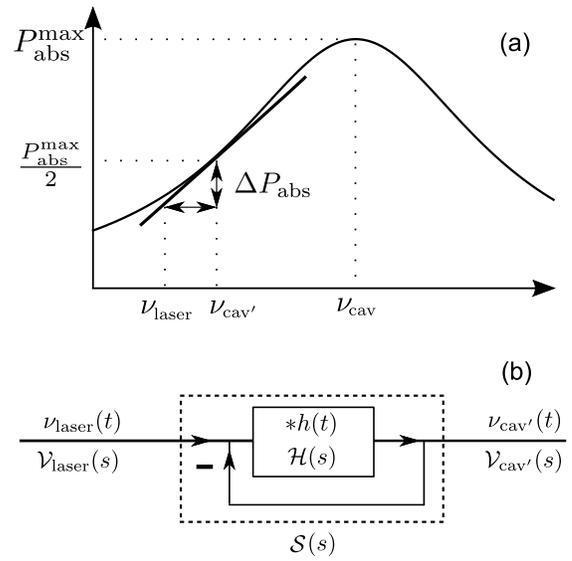}
  	\caption{Thermal self-locking mechanism in cavities. {\bf a} Part of the laser light that couples into a cavity gets absorbed and results in a thermal response and a change in the cavity resonance. This in turn affects the amount of light that enters the cavity, which depends on the relative detuning between the laser and the cavity. {\bf b} By linearizing the closed-loop response around a stable locking point half-way up the Lorentzian cavity resonance, the behavior of the self-locked cavity can be modeled as a linear time-invariant system.}{\label{fig:selflock}}
\end{figure}
%%%%%%%%%%%%%%%%%%%%%%%%%%%%%%%%%%%%%%%%%%%%%%%%%%%%%%%%%%%%%%%%
\subsubsection*{Thermal self-locking in FFPCs} 
Thermally induced nonlinearities in optical cavities are understood to arise from thermal cavity response to the absorption of light that circulates in the optical resonator. Previously, the dynamical behavior and stability conditions resulting from cavity self-heating have been studied in detail for whispering gallery micro-resonators~\cite{Carmon2004}. Thermo-optical bistability has furthermore been used to measure the absorption of mirror coatings of traditional substrates~\cite{An1997} and fiber cavity mirrors~\cite{Hunger2010}. 
Our goal in investigating the thermal effects for FFPCs is to understand and optimize the dynamics of the system in the self-stable regime~\cite{Carmon2004}. Self-locking conditions, where the system automatically corrects for drifts and deviations, are important for operating the rigid cavity with strong intra-cavity fields, such as they are needed for the optical trapping of particles in neutral atom CQED and cavity optomechanics~\cite{Asenbaum2013}. 
%%%%%%%%%%%%%%%%%%%%%%%%%%%%%%%%%%%%%%%%%%%%%%%%%%%%%%%%%%%%%%%%
\begin{figure*}[ht]
    \includegraphics[width=0.9\textwidth]{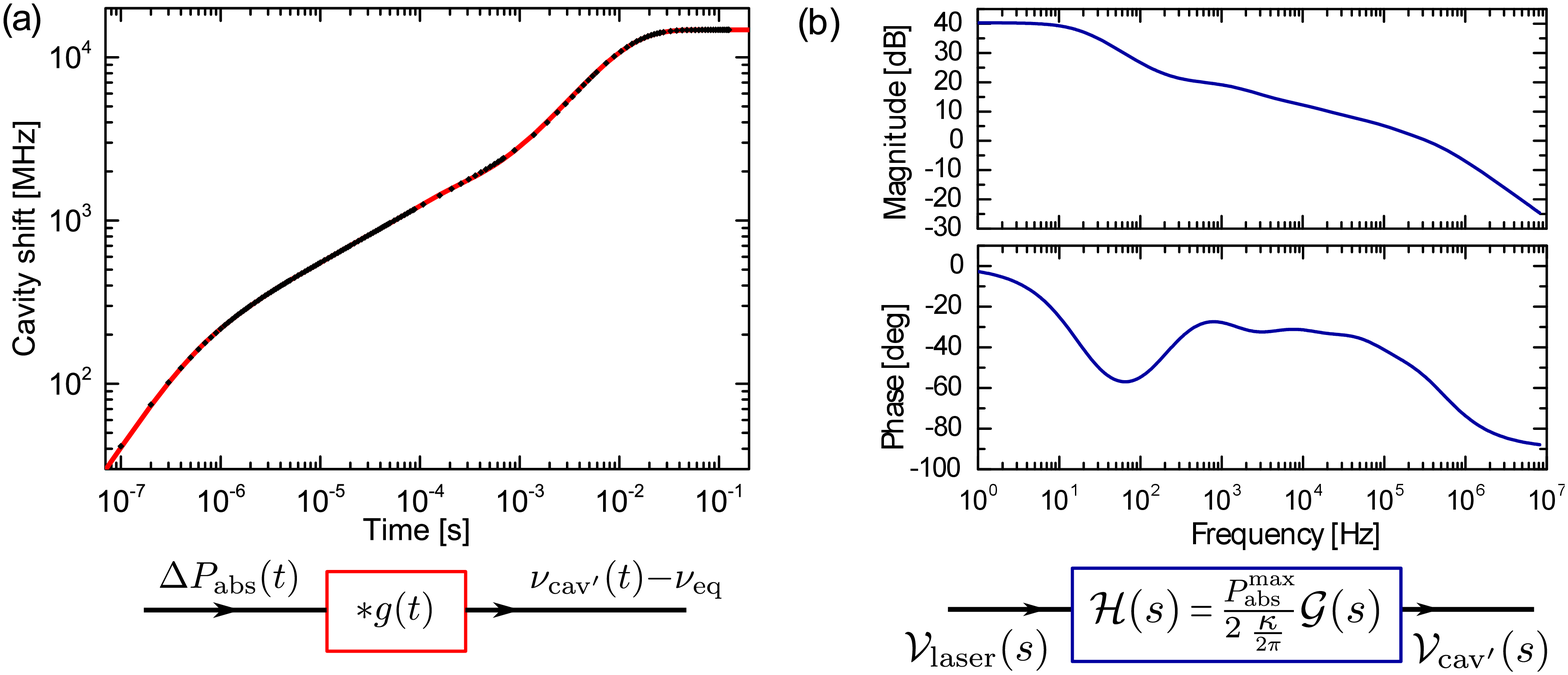}
  	\caption{Theoretical open-loop response of the rigid fiber cavity. {\bf a} Step response of the cavity resonance frequency for a step change in the absorbed power of $\Delta P_{\rm{abs}} = 1\,$mW obtained from a finite element simulation. The numerical step response results are fitted with a curve that is the sum of five exponential functions and an offset. The impulse response $g(t)$ is the time derivative of the step response. {\bf b} Bode diagrams of the transfer function $\mathcal{H}(s)$, which follows from the Laplace transform $\mathcal{G}(s)$ of $g(t)$ and appropriate normalization, for $P_{\rm{abs}}^{\rm{max}} = 1\,$mW and $\kappa = 2\pi\cdot65\,$MHz.}{\label{fig:response}}
\end{figure*}
%%%%%%%%%%%%%%%%%%%%%%%%%%%%%%%%%%%%%%%%%%%%%%%%%%%%%%%%%%%%%%%%

In the following we consider the self-locking scenario sketched in Fig.~\ref{fig:selflock}a. When a laser light field with power $P_{\rm{in}}$ and frequency $\nu_{\rm{laser}}$ impinges onto a FFPC, part of the power that builds up in the cavity is absorbed by the mirror coatings and acts as a localized heat input to the cavity assembly. The absorbed power in each mirror is described by
\begin{align}
\label{eq:powerabsorbed}
P_{\rm{abs}}(\nu_{\rm{laser}}, \nu_{\rm{cav}}) &=  \tfrac{4\epsilon_{\rm{in}} \cdot \mathcal{A} \cdot \mathcal{T}_{\rm{in}} \cdot P_{\rm{in}}}{(\mathcal{L}_{\rm{in}}+\mathcal{L}_{\rm{out}}+\mathcal{T}_{\rm{in}}+\mathcal{T}_{\rm{out}})^2} \, \mathcalboondox{L}(\nu_{\rm{laser}},\nu_{\rm{cav}})\\ \nonumber
&= P^{\rm{max}}_{\rm{abs}} \, \mathcalboondox{L}(\nu_{\rm{laser}},\nu_{\rm{cav}})
\end{align}
where $\mathcal{A}$ is the mirror absorption loss and $\mathcalboondox{L}(\nu_{\rm{laser}}, \nu_{\rm{cav}})\!=\! 1/(1\!+\!(\tfrac{2\pi(\nu_{\rm{laser}}\!-\!\nu_{\rm{cav}})}{\kappa})^2)$ denotes the Lorentzian cavity lineshape.
According to Eq.~\ref{eq:powerabsorbed}, variations in the instantaneous detuning of the laser frequency from the cavity resonance directly translate into variations in the absorbed laser power. If the corresponding change in temperature of the FFPC system gives rise to a modification of cavity length (and therefore cavity resonance frequency) that at least partially compensates the initial detuning, then the cavity is located in the self-locking regime. 

For the investigation of the self-locking dynamics we adjust the frequency of the incident laser to operate the cavity in an equilibrium point on the slope of the resonance at $\nu_{\rm{laser}}\!=\!\nu_{\rm{cav'}}\!=\! \nu_{\rm{cav}}-\tfrac{\kappa}{2\pi}$ where the absorbed power is $P_{\rm{abs}}^{\rm{max}}/2$. By linearizing Eq.~\ref{eq:powerabsorbed} around the equilibrium point we obtain $\Delta P_{\rm{abs}}(\nu_{\rm{laser}}, \nu_{\rm{cav}}) \! \approx\! \frac{P_{\rm{abs}}^{\rm{max}}}{2}\cdot \frac{2\pi(\nu_{\rm{laser}} - \nu_{\rm{cav'}})}{\kappa}$ (see Fig.~\ref{fig:selflock}a). In this way we can model cavity self-locking as a linear time-invariant system and analyze the dynamics using the formalism and standard tools of control theory. The behavior of the self-locked cavity around its equilibrium point $\nu_{\rm{eq}}$ can then be viewed as a closed-loop feedback system and can be described in the time domain by         
\begin{align}
\label{eq:closedloop}
\nu_{\rm{cav'}}(t)-\nu_{\rm{eq}} &= \Big((\nu_{\rm{laser}}-\nu_{\rm{cav'}}) * h\Big)(t) \\ \nonumber
&= \int_0 ^t\! \Big(\nu_{\rm{laser}}(t')-\nu_{\rm{cav'}}(t')\Big) h(t-t')dt',
\end{align}
where $h(t)$ is the open-loop impulse response of the system (see Fig.~\ref{fig:selflock}b). 
The equivalent description in the frequency domain is obtained from the Laplace transform of Eq.~\ref{eq:closedloop} and reads 
\begin{equation}
\mathcal{V}_{\rm{cav'}}(s) = \mathcal{S}(s)\cdot\mathcal{V}_{\rm{laser}}(s) = \frac{\mathcal{H}(s)}{1+\mathcal{H}(s)}\cdot\mathcal{V}_{\rm{laser}}(s),
\label{eq:closedlooptransformed}
\end{equation}
where $\mathcal{H}(s)$ and $\mathcal{S}(s)$ are the open-loop and closed-loop transfer functions of the system, respectively. 

\begin{figure}[ht]
    \includegraphics[width=1\columnwidth]{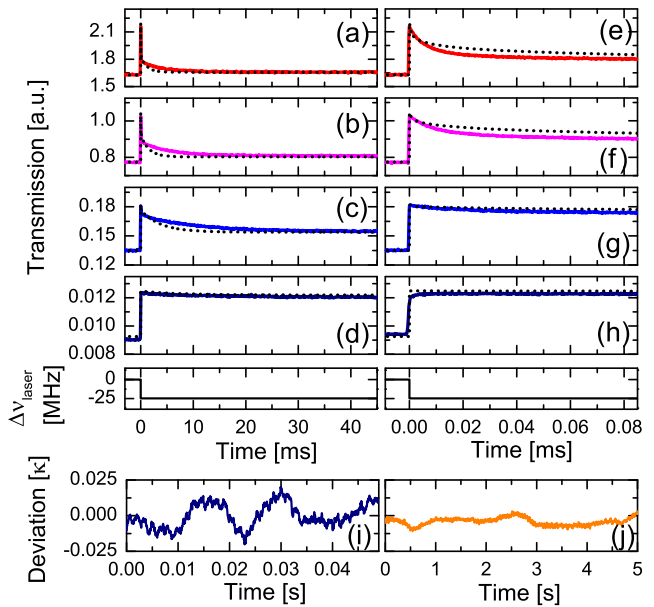} 
  	\caption{Experimental (solid line, average over 32 repetitions) and simulated (dotted line) step responses of the self-locked cavity to a step change of the laser frequency for incident laser powers $P_{\rm{in}} = \left\{ 5.5,2.5,0.5,0.03\right\}\,$mW at the ({\bf a-d}) millisecond  and ({\bf e-h}) microsecond  timescale. Stability of the rigid fiber cavity when coupled with laser light of $0.03\,$mW ({\bf i}) and  $62\,$mW ({\bf j}) incident power.}{\label{fig:closedloop}}
\end{figure}

In the experiment we can characterize the self-locked cavity system by probing its (closed-loop) response to a step in the laser frequency $\nu_{\rm{laser}}$. For this purpose a $20\,$Hz square-wave modulation with frequency deviation of 24.8 MHz is applied to the laser frequency while its transmission through the FFPC on the approximately linear slope of the resonance is monitored as a function of time. The results for different incident laser powers $P_{\rm{in}}$ are shown in Fig.~\ref{fig:closedloop}a-h. It is evident that for increasing laser powers the stiffness of the self-lock increases, leading to a faster response of the lock and to a smaller relative error in the steady state.   

In order to theoretically model the observed self-locking behavior of the FFPC we assume that the dominant mechanism for the thermal response is the change of cavity length due to the thermal expansion of the fiber coating and substrate~\cite{An1997,Hunger2010}. A finite element simulation is used to calculate the temperature distribution in the fiber as a function of time after a step heat input with the spatial profile of the cavity mode is applied to the front surface of the SiO$_2$-Ta$_2$O$_5$ multilayer coating. We treat the coating as a homogeneous but anisotropic layer of $L_{\rm{coat}}=4\,\mu$m thickness, considering averaged material properties of SiO$_2$ and Ta$_2$O$_5$~\cite{Crooks2006}. All boundaries of the optical fiber protruding from the ferrule by $L_{\rm{fiber}}=126\,\mu$m are considered to be isolating, except for the back surface, which is assumed to remain at a constant temperature due to its thermal contact with the ferrule. The thermal conductivity of the fiber and the fiber coating material are considered to be the only mechanisms of heat transfer in the system. From the time-resolved distribution of temperature increases $\Delta T(z,t)$ along the axis of the optical fiber, the total length change of the fiber mirror  
\begin{equation}
\Delta L_{\rm{mirror}}(t) \!=\! c_{\rm{coat}}\!\! \int \limits^{L_{\rm{coat}}}_0 \!\! \Delta T(z,t) dz + c_{\rm{fiber}} \!\! \int \limits^{L_{\rm{fiber}}}_{L_{\rm{coat}}} \!\! \Delta T(z,t) dz
\label{eq:}
\end{equation}
follows, where $c_{\rm{coat}}$ and $c_{\rm{fiber}}$ are the coefficients of linear thermal expansion of the coating and the fiber material, respectively. The shift of the cavity resonance is finally given by  
\begin{equation}
\Delta \nu_{\rm{cav'}}(t)\! =\! - \frac{\nu_{\rm{cav'}}}{L_{\rm{cav}}}\Delta L_{\rm{cav}}(t)\! =\! + \frac{\nu_{\rm{cav'}}}{L_{\rm{cav}}} \cdot2\,\Delta L_{\rm{mirror}}(t). 
\label{eq:}
\end{equation}

The theoretical step response to a simulated step input of $\Delta P_{\rm{abs}} = 1\,$mW is shown in Fig.~\ref{fig:response}a. From the multi-exponential fit to the step response we obtain the impulse response function $g(t)$ by taking the time derivative. The Laplace transform of the appropriately normalized impulse response provides the desired open-loop transfer function $H(s)$ for which the Bode plots are shown in Fig.~\ref{fig:response}b with values $P_{\rm{abs}}^{\rm{max}} = 1\,$mW and $\kappa = 2\pi\cdot65\,$MHz. 

For comparison with the experimental data in Fig.~\ref{fig:closedloop}a-h we compute the closed-loop response to a laser frequency step change for different input powers ($P_{\rm{in}} = \left\{ 5.5,2.5,0.5,0.03\right\}\,$mW, $\pm 10\%$ uncertainty) using Eqs.~\ref{eq:transmission}, \ref{eq:powerabsorbed} and \ref{eq:closedlooptransformed}. Considering the stated values of $\mathcal{F}$, $\mathcal{T}_{\rm{in,850\,\rm{nm}}}$, $P_{\rm{in}}$ and the estimate $\epsilon_{\rm{in}}\approx 0.55$, we find agreement between experimental data and the results of the theoretical model, shown in Fig.~\ref{fig:closedloop}a-h, for a mirror absorption loss of $\mathcal{A}\sim10\,$ppm, which is consistent with the theoretically expected value. We attribute residual discrepancies between the theoretical and experimental curves to the simplified treatment of heat diffusion and thermal expansion in the coating layer and at the glued ferrule-fiber boundary, which could misestimate the theoretical cavity response at the short and long timescales, respectively. Potential contributions to the cavity length change from thermally induced stress at the coating-fiber interface would also merit further investigations but are beyond the scope of this article.   
     
Due to the assumption of instantaneous thermal expansion, the self-locking model predicts stable locking behavior even for arbitrarily high gain ($\propto P_{\rm{in}}$). Experimentally, we have observed increasingly stable self-locking behavior up to the highest tested incident laser power of $62\,$mW corresponding to an intracavity power of $\sim90\,$W (see Fig.~\ref{fig:closedloop}j).
%%%%%%%%%%%%%%%%%%%%%%%%%%%%%%%%%%%%%%%%%%%%%%%%%%%%%%%%%%%%%%%%
%%%%%%%%%%%%%%%%%%%%%%%%%%%%%%%%%%%%%%%%%%%%%%%%%%%%%%%%%%%%%%%%								 
\section{Theory of mode matching and reflective line shapes in FFPCs}
\label{sec:theorymodematching}
The following discussion is separated into two parts.\\
First, we formulate the reflective mode matching problem for free-space coupled Fabry-Perot cavities and FFPCs, and we show that the connection between the cavity mode matching and the observable cavity reflection signals fundamentally differs for the two cases. \\
In the second part, the model for the FFPCs is evaluated in terms of the fiber cavity mirror parameters and alignment degrees of freedom. The predictions of the model are compared with experimental results for a strongly overcoupled FFPC and the implications for the optimal alignment of FFPCs are discussed. Details of the calculation and the computational routines are provided in a script~\footnote{Wolfram Mathematica\textsuperscript{\textregistered}} in the Supplemental Material~\cite{Note3}.   

\subsubsection*{Notation}
\begin{figure*}
    \includegraphics[width=0.900\textwidth]{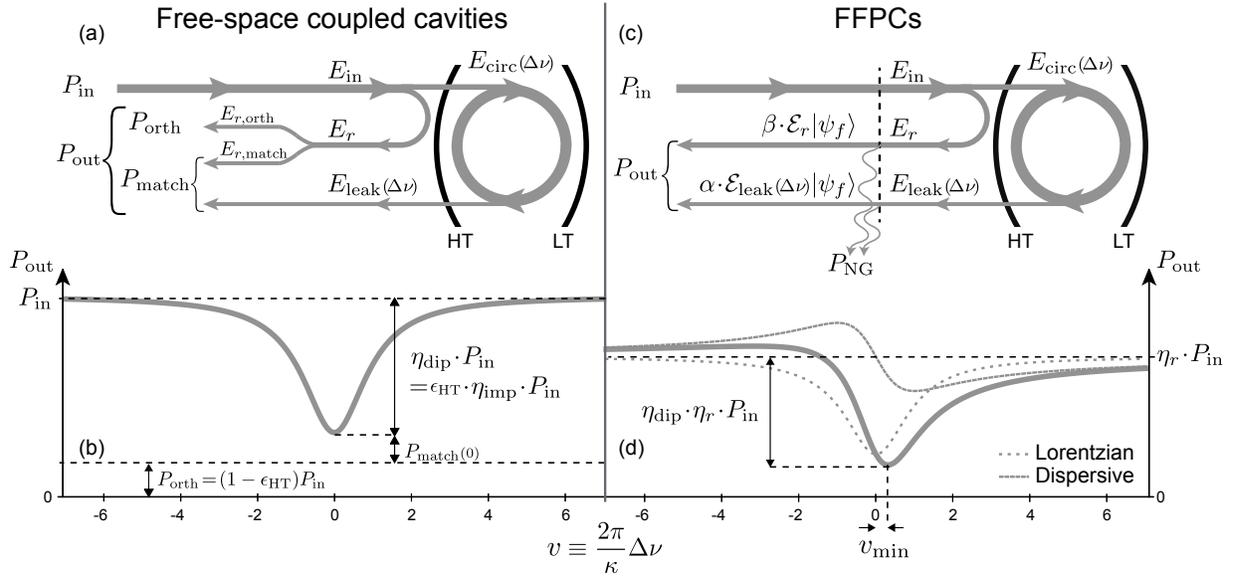}
	\caption{Illustration of the contributions to the reflective cavity line shapes. For free-space coupled cavities, the interference of the spatially mode matched part of the directly reflected field $E_{r}$ interferes with the cavity leakage field $E_{\rm{leak}}$ ({\bf a}) and gives rise to the Lorentzian dip in the cavity reflected power at resonance ({\bf b}). For FFPCs, the mode filtering of the reflected light by the single mode fiber implies that the two fields can fully interfere ({\bf c}) and the depth of the asymmetric line shape is no longer directly related to the spatial mode efficiency $\epsilon_{\HT}$ ({\bf d}). See main text for details.}{\label{fig:intuitive}}
\end{figure*}

To keep our notation compact, we write all electric fields involved in the formulation of the problem (see Fig.~\ref{fig:intuitive}) as 
\begin{equation}
E_i = \mathcal{E}_i \cdot \ket{\psi_i},
\label{eq:electricfields}
\end{equation}
where the Dirac notation is used for the description of the spatial mode components~\cite{Joyce1984} and $\mathcal{E}_i$ contains the complex amplitude of the field, including the time dependence $e^{i\omega t}$. The overlap integrals of normalized spatial mode functions can thus be basis-independently expressed as $\braket{\psi_i}{\psi_j}$.     

For the description of the mode matching geometry shown in Fig.~\ref{fig:modematching} we introduce 4 spatial modes: the forward and backward propagating modes in the single mode fiber $\ket{\psi^\pm_{{f}}}$, the spatial mode $\ket{\psi_{r}}$ corresponding to the fiber mirror reflection of the forward propagating fiber mode $\ket{\psi^+_{f}}$, and the forward and backward propagating cavity modes $\ket{\psi_{\rm{cav}}^{\pm}}$.\\
We name their mutual overlap integrals 
\begin{align}
\label{eq:namedoverlaps}
&\braket{\psi^+_{\rm{cav}}}{\psi^+_f}\! =\! \alpha \nonumber \\ 
&\braket{\psi^-_{\rm{cav}}}{\psi_r }\!=\! \bra{\psi^+_{\rm{cav}}}R^{\dagger}R\ket{\psi^+_f}\!=\!\braket{\psi^+_{\rm{cav}}}{\psi^+_f} \!=\! \alpha  \nonumber \\
&\braket{\psi^-_{\rm{cav}}}{\psi^-_f }\!=\! (\braket{\psi^+_{\rm{cav}}}{\psi^+_f})^* \!=\! \alpha^*  \nonumber \\ 
&\braket{\psi^-_{f}}{\psi_r}  \!=\! \beta,  
\end{align}
with the overlap amplitudes $\alpha,\beta \! \in \!\mathbb{C}$. The mode matching efficiency from the fiber mode to the cavity is then given by~\cite{Joyce1984}
\begin{equation}
\epsilon_\HT = |\braket{\psi^+_{\rm{cav}}}{\psi^+_f}|^2 = |\alpha|^2.
\label{eq:modematch}
\end{equation}
Furthermore, in Eq.~\ref{eq:namedoverlaps} we utilize the fact that reflection (from the incoupling mirror) is a unitary transformation ($R^{\dagger}R=\mathds{1}$) and that overlap amplitudes are conjugated when the directions of propagation of both modes are changed~\footnote{This is evident for the overlap integral of Gaussian beams,
  where for the complex spatial mode functions the relation $\psi_{i,\vec{k}}(x,y,z)= \psi_{i,-\vec{k}}^*(x,y,z)$ holds and hence $\int_{-\infty}^{\infty}\int_{-\infty}^{\infty}\psi^*_{1,\vec{k}_1}(x,y,z)\psi_{2,\vec{k}_2}(x,y,z)\, dxdy = (\int_{-\infty}^{\infty}\int_{-\infty}^{\infty}\psi^*_{1,-\vec{k}_1}(x,y,z)\psi_{2,-\vec{k}_2}(x,y,z)\, dxdy)^*$ (see Supplemental Material~\cite{Note3} for details).}.  
%%%%%%%%%%%%%%%%%%%%%%%%%%%%%%%%%%%%%%%%%%%%%%%%%%%%%%%%%%%%%%%%
%%%%%%%%%%%%%%%%%%%%%%%%%%%%%%%%%%%%%%%%%%%%%%%%%%%%%%%%%%%%%%%%
\subsection{Formulation of the mode matching problem}
\subsubsection*{Free-space coupled cavities}
We start by revisiting the reflective mode matching problem for the well known case of free-space coupled cavities~\cite{Hood2001}. Fig.~\ref{fig:intuitive}a illustrates that for the incident light field 
\begin{equation}
E_{\rm{in}} = \mathcal{E}_{\rm{in}} \ket{\psi^+_f} 
\label{eq:incident}
\end{equation}
the reflected signal on a photo detector can be understood from the interference of the two light fields $E_{\rm{leak}}(\Delta \nu)$ and $E_r$, where $\Delta \nu=\nu_{\rm{laser}} - \nu_{\rm{cav}}$.\\
The transmission of the detuning-dependent intra-cavity field 
$E_{\rm{circ}}(\Delta \nu) = \mathcal{E}_{\rm{in}}\frac{i\,t_{\rm{HT}} \braket{\psi^+_{\rm{cav}}}{\psi^+_f}}{1-r_{\rm{HT}} r_{\rm{LT}} e^{i \phi}} \ket{\psi_{\rm{cav}}^+}$
through the incoupling mirror gives rise to the leakage field
\begin{align}
\label{eq:leak}
E_{\rm{leak}} (\Delta \nu) &=  i\,t_\HT \cdot \mathcal{E}_{\rm{circ}} \cdot r_\LT e^{i\phi} \ket{\psi_{\rm{cav}}^-} \\ \nonumber
&= - \mathcal{E}_{\rm{in}}\frac{t^2_\HT \, r_\LT \,e^{i \phi}}{1-r_\HT \,r_\LT \,e^{i \phi}} \braket{\psi^+_{\rm{cav}}}{\psi^+_f} \ket{\psi_{\rm{cav}}^-} \\ \nonumber
&= -\mathcal{E}_{\rm{in}} \cdot \zeta(\Delta\nu) \cdot \braket{\psi^+_{\rm{cav}}}{\psi^+_f} \ket{\psi_{\rm{cav}}^-},
\end{align}
where a phase factor $\tfrac{\pi}{2}$ upon mirror transmission and the round-trip phase $\phi\!=2\pi \tfrac{\Delta \nu\cdot 2L_{\rm{cav}}}{c}\!=\!2\pi\frac{\Delta \nu}{\Delta \nu_{\rm{FSR}}}$ have been considered \cite{Siegman1986}. $r_i$, $t_i$ are reflection and transmission coefficient of cavity mirror $i$. $c$ and $\Delta \nu_{\rm{FSR}}$ denote the speed of light and the cavity free spectral range, respectively. Additionally, for later convenience, the complex quantity $\zeta(\Delta \nu)$ is introduced in Eq.~\ref{eq:leak} to describe the leakage field's dependence on mirror properties and the cavity length.
%%%%%%%%%%%%%%%%%%%%%%%%%%%%%%%%%%%%%%%%%%%%%%%%%%%%%%%%%%%%%%%%
\begin{figure}[hb]
    \includegraphics[width=0.80\columnwidth]{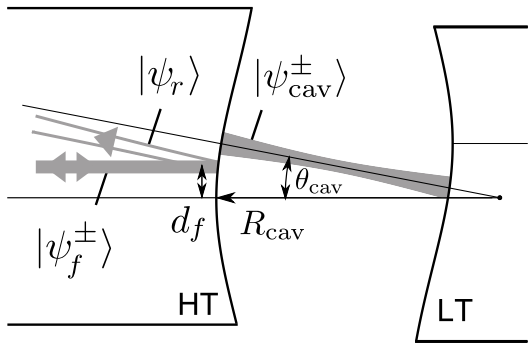}
	\caption{Geometry of the FFPC mode matching problem. The relative alignment of the forward and backward propagating fiber ($\ket{\psi^\pm_{{f}}}$) and cavity ($\ket{\psi^\pm_{\rm{cav}}}$) modes and the mode ($\ket{\psi_{r}}$) corresponding to the fiber mirror reflection of $\ket{\psi^+_{{f}}}$ is shown along one spatial dimension. An equivalent description holds for the other transverse direction.}{\label{fig:modematching}}
\end{figure}
%%%%%%%%%%%%%%%%%%%%%%%%%%%%%%%%%%%%%%%%%%%%%%%%%%%%%%%%%%%%%%%%

The field directly reflected from the incoupling mirror
\begin{equation}
E_{r} = r_\HT \cdot\mathcal{E}_{\rm{in}} \ket{\psi_{r}} \simeq \mathcal{E}_{\rm{in}} \ket{\psi_{r}}, 
\label{eq:}
\end{equation} 
can be decomposed into a part 
\begin{equation}
E_{r,\rm{match}} = \Big(\bra{\psi^-_{\rm{cav}}} E_{r}\Big) \ket{\psi^-_{\rm{cav}}} = \mathcal{E}_{\rm{in}} \braket{\psi^-_{\rm{cav}}}{\psi_{r}} \ket{\psi^-_{\rm{cav}}}
\label{eq:}
\end{equation}
that is mode matched to the cavity and therefore interferes with $E_{\rm{leak}} (\Delta \nu)$, and a part    
\begin{equation}
E_{r,\rm{orth}} = E_r - E_{r,\rm{match}} = \mathcal{E}_{\rm{in}} \Big(\ket{\psi_r}-\braket{\psi^-_{\rm{cav}}}{\psi_{r}} \ket{\psi^-_{\rm{cav}}}\Big)
\label{eq:orth}
\end{equation} 
that is orthogonal to the cavity mode and by definition does not interfere. \\
The total power recorded at the detector is given by
\begin{align}
\label{eq:powertraditional}
P_{\rm{out}} =& P_{\rm{orth}} +P_{\rm{match}}(\Delta \nu) \\ \nonumber
=& E^\dagger_{r,\rm{orth}}E_{r,\rm{orth}} \! \\ \nonumber
&+\! \big(E_{r,\rm{match}} \!+\! E_{\rm{leak}}(\Delta \nu)\big)^\dagger \big(E_{r,\rm{match}}\!+\! E_{\rm{leak}}(\Delta \nu)\big)  \\ \nonumber
=& P_{\rm{in}}  \big(1-\epsilon_{\HT}\big) + P_{\rm{in}} \Big(\epsilon_{\HT} \big|1-\zeta(\Delta\nu)\big|^2\Big).
\end{align}  
The Taylor expansion of Eq.~\ref{eq:powertraditional} around the resonance $\nu_{\rm{cav}}$ (for $\Delta \nu/\Delta \nu_{\rm{FSR}} \ll 1$) yields the familiar result of Lorentzian dips in the reflected power at the cavity resonances 
\begin{align}
\label{eq:resonances}
\frac{P_{\rm{out}}(v)}{P_{\rm{in}}} &= 1- \eta_{\rm{dip}} \frac{1}{1+v^2},
\end{align}
where $v=2\pi(\nu_{\rm{laser}}-\nu_{\rm{cav}})/\kappa$ is the normalized detuning. The dip amplitude
\begin{align}
\label{eq:dip}
\eta_{\rm{dip}}&=\epsilon_{\HT} \cdot  \big(1-\big|1-\zeta(0)\big|^2\big) \\ \nonumber
 &= \epsilon_{\HT} \cdot  \Bigg(1-\frac{(\mathcal{T}_\HT-\mathcal{T}_\LT-\mathcal{L}_\HT-\mathcal{L}_\LT)^2}{\mathcal{L}_{\rm{sum}}^2}\Bigg) \\ \nonumber &= \epsilon_{\HT} \cdot \eta_{\rm{imp}}\\
 \rm{with}& \quad \mathcal{L}_{\rm{sum}} = \mathcal{T}_\HT+\mathcal{T}_\LT+\mathcal{L}_\HT+\mathcal{L}_\LT \nonumber
\end{align}
is directly proportional to the spatial mode matching $\epsilon_{\HT}$, which can thus be easily determined once the mirror parameters are independently measured. 
According to Eq.~\ref{eq:dip} the depth of the reflective dip $\eta_{\rm{dip}}$ is maximum for cavity mirrors that satisfy the impedance matching condition $(\mathcal{T}_\HT = \mathcal{T}_\LT+\mathcal{L}_\HT+\mathcal{L}_\LT; \eta_{\rm{imp}} = 1)$ and decreases symmetrically for equally overcoupled and undercoupled cavities. 
\subsubsection*{FFPCs and spatially mode filtered cavities}
The reflective mode matching for FFPCs can be imagined as the case of a free-space coupled cavity followed by spatial mode filtering of the returning light by the single mode fiber (see Fig.~\ref{fig:intuitive}b). Only light that couples back into the fiber mode is guided back to the detector, whereas the light entering the cladding modes is lost and does not reach the detector.  

The light field that is guided back in the fiber mode 
\begin{align}
\label{eq:eoutffpc}
E_{\rm{out}}(\Delta\nu) &= \Big(\bra{\psi^-_f}E_r + \bra{\psi^-_f}E_{\rm{leak}}(\Delta\nu)\Big) \ket{\psi^-_f}\\ \nonumber
= \mathcal{E}_{\rm{in}} &\bigg(\braket{\psi^-_f}{\psi_r} + \zeta(\Delta\nu) \braket{\psi^+_{\rm{cav}}}{\psi^+_f} \braket{\psi^-_f}{\psi^-_{\rm{cav}}}\bigg)\ket{\psi^-_f}
\end{align}
thus results in the detected power  
\begin{align}
P_{\rm{out}} &= E^\dagger_{\rm{out}}E_{\rm{out}}  \\ \nonumber
&= P_{\rm{in}} \big|\beta-\alpha^2 \cdot \zeta(\Delta\nu)\big|^2  \\ \nonumber
&= P_{\rm{in}} \bigg| \frac{1}{\epsilon_{\HT}}\Big(\beta \cdot (\alpha^2)^*-\epsilon_{\HT}^2 \cdot \zeta(\Delta\nu)\Big)\bigg|^2.
\label{eq:poutffpc}
\end{align}  
The Taylor approximated ($\Delta \nu/\Delta \nu_{\rm{FSR}} \ll 1$) solution
\begin{equation}\label{eq:mainresult}
\frac{P_{\rm{out}}(v)}{P_{\rm{in}}}=\eta_{r} -\eta_{\scriptscriptstyle {\mathcalboondox{L}}}\left(\frac{1}{1+v^2}-\mathcalboondox{A}\frac{v}{1+v^2}\right)
\end{equation}
features an asymmetric dip in the reflected power at the cavity resonance that consists of a Lorentzian of amplitude 
\begin{equation}
\eta_{\scriptscriptstyle {\mathcalboondox{L}}} = \frac{4\,\mathcal{T}_\HT}{\mathcal{L}_{\rm{sum}}}\left(\operatorname{Re}\big[\beta \cdot(\alpha^2)^*\big]-\epsilon_{\HT}^2\,\frac{\mathcal{T}_\HT}{\mathcal{L}_{\rm{sum}}}\right)
\label{eq:FFPCdip}
\end{equation}
and its corresponding dispersive curve with relative amplitude 
\begin{equation}
\mathcalboondox{A}=\frac{\operatorname{Im}\big[\beta \cdot(\alpha^2)^*\big]}{\operatorname{Re}\!\big[\beta\cdot(\alpha^2)^*\big]-\epsilon^2_{\HT} \frac{\mathcal{T}_\HT}{\mathcal{L}_{\rm{sum}}}},
\label{eq:asymmetryfactor}
\end{equation} 
where $\operatorname{Re}\big[...]$ and $\operatorname{Im}\big[...\big]$ denote the real and imaginary part.  

The comparison of Eq.~\ref{eq:mainresult} with Eq.~\ref{eq:resonances} in Fig.~\ref{fig:intuitive} illustrates that spatial mode filtering fundamentally changes the reflective spectral line shape. 
In the off-resonant case, the divergence and deflection upon reflection on a curved and decentered fiber mirror lead to power losses $P_{\rm{NG}}$ into non-guiding fiber cladding modes that reduce the fraction of incident light that reaches the detector by $\eta_r = |\beta|^2$. The depth $\eta_{\rm{dip}}$ of the asymmetric FFPC dip is no longer limited by $\epsilon_{\HT}$, is not maximum for the impedance matched condition $(\mathcal{T}_\HT = \mathcal{T}_\LT+\mathcal{L}_\HT+\mathcal{L}_\LT)$ and responds differently to overcoupling or undercoupling. In particular for significantly overcoupled cavities, the state $\eta_{\rm{dip}}=1$ can always be reached for sufficient misalignment, i.e. reduction of $\epsilon_{\rm{HT}}$.  

No differences arise in the line shape of the circulating power and the transmission of FFPCs, which are Lorentzian as in the case of free-space coupled cavities. The minimum of the reflective power $\frac{P_{\rm{out}}(v_{\rm{min}})}{P_{\rm{in}}} = \eta_r \big(1-\eta_{\rm{dip}}\big)$ is hence shifted relative to the (transmittive) cavity resonance frequency by $v_{\rm{min}}=\frac{1}{\mathcalboondox{A}}\left(1-\sqrt{1+\mathcalboondox{A}^2}\right)$.

\subsection{Optimal FFPC alignment}
The geometry of FFPC alignments is sketched in Fig.~\ref{fig:modematching}. By factoring the overlap integrals along two orthogonal directions $\braket{\psi_i}{\psi_j} = \braket{\psi_{i,x}}{\psi_{j,x}}\braket{\psi_{i,y}}{\psi_{j,y}}$, the problem can be analyzed by considering two individual one-dimensional alignments~\cite{Joyce1984}. 
Optimal alignment for a one-sided cavity corresponds to maximizing the spatial mode matching efficiency $\epsilon_\HT=|\alpha|^2$ from the cavity mode to the HT-fiber. This is achieved by adjusting the cavity angles $\theta_{{\rm{cav}},x}$ and $\theta_{{\rm{cav}},y}$ using the transversal displacement or the relative inclination of the fiber mirrors. 

%%%%%%%%%%%%%%%%%%%%%%%%%%%%%%%%%%%%%%%%%%%%%%%%%%%%%%%%%%%%%%%
\begin{figure}[t]
\includegraphics[width=1.00\columnwidth]{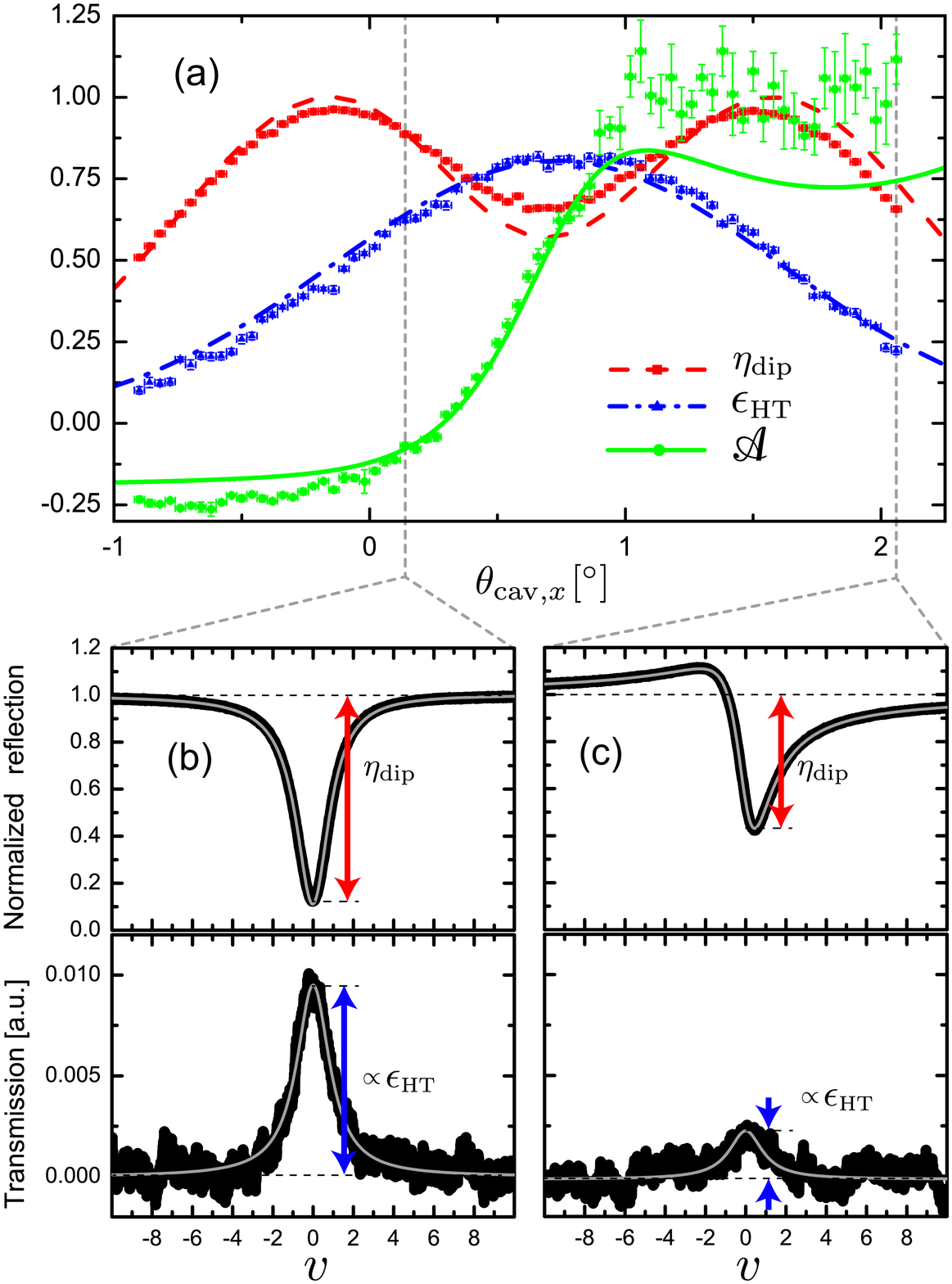}
\caption{Experimental verification of the FFPC alignment model. {\bf a} Spatial mode matching efficiency $\epsilon_{\rm{HT}}$, depth of dip in reflection  $\eta_{\rm{dip}}$ and asymmetry parameter $\mathcalboondox{A}$ as a function of the cavity alignment. The theoretical model (lines) agrees well with the experimental data points for the strongly overcoupled FFPC. The error bars in y-direction denote the one-sigma ``standard error of the mean'' from 10 identical measurements. x-errors denote the estimated precision of the rotation stage. The optimal cavity alignment occurs for a cavity mode angle of $0.8^{\circ}$ due to the significant decentration of the mirror fiber. The experimentally observed maximum of $\eta_{\rm{dip}}$ is reduced compared to the theoretically predicted value $\eta_{\rm{dip, max}}=1$ due to the background light from reflections along the optical path. Reflective and transmittive line shapes are shown for the cavity angles $\sim0^{\circ}$ ({\bf{b}}) and $\sim2^{\circ}$ ({\bf{c}}).} {\label{fig:opencavity}}
\end{figure}
%%%%%%%%%%%%%%%%%%%%%%%%%%%%%%%%%%%%%%%%%%%%%%%%%%%%%%%%%%%%%%%

For the theoretical modeling of the alignment-dependent mode matching and reflective line shapes with Eqs.~\ref{eq:modematch} and~\ref{eq:mainresult} the mode overlap amplitudes $\alpha$ and $\beta$ need to be evaluated in terms of the cavity mirror parameter (radii of curvature $R_\HT$, $R_\LT$, mirror decentrations $d_{f,x}$, $d_{f,y}$ and fiber mode field radius $w_f$) and the alignment parameters (cavity length $L_{\rm{cav}}$ and cavity alignment angles $\theta_{{\rm{cav}},x}$ and $\theta_{{\rm{cav}},y}$).
We find analytic expressions for $\alpha$ and $\beta$ by performing the overlap integration of the modes $\ket{\psi^\pm_f}$, $\ket{\psi^\pm_{\rm{cav}}}$ and $\ket{\psi_r}$ represented as inclined and displaced Gaussian beams, from where we finally obtain analytic expression for $\mathcalboondox{A}$, $\eta_{\rm{dip}}$ and $\epsilon_{\HT}$ (for details of the calculation see the Supplemental Material~\cite{Note3}).  

To experimentally confirm the calculation we perform measurements on a strongly overcoupled hybrid cavity that is formed by the combination of a HT fiber mirror and a macroscopic mirror substrate with ultra-low transmission ($\mathcal{T}_{\rm{ULT,850\,\rm{nm}}}=0.9\,$ppm, $\mathcal{L}_{\rm{ULT,850\,\rm{nm}}}=4\,$ppm, $R_{\rm{ULT}} = 50\,000\,\mu$m). We record the reflective as well as the transmitted line shape as a function of the inclination angle $\theta_m$ of the mirror substrate. Due to the gimbal mounting of the nearly flat ($R_{\rm{ULT}} \gg R_{\rm{HT}}$) ULT mirror we can make the approximations $\theta_m \simeq \theta_{{\rm{cav}},x}$ and $L_{\rm{cav}}(\theta_{{\rm{cav}},x}) \simeq L_{\rm{cav}}(0)$. By fitting Eq.~\ref{eq:mainresult} to the the reflective line shapes we obtain $\mathcalboondox{A}$ and $\eta_{\rm{dip}}$ (see Fig.~\ref{fig:opencavity}b and c). Moreover, since the mode matching efficiency on the ULT mirror side can be assumed as independent of the alignment ($\epsilon_{\rm{ULT}}\approx 1$), it follows from Eq.~\ref{eq:transmission} that the fiber to cavity mode matching efficiency $\epsilon_{\rm{HT}}(\theta_{{\rm{cav}},x}) = C_T \cdot P_{\rm{T,max}}(\theta_{{\rm{cav}},x})$ is proportional to the resonant cavity transmission signal $P_{\rm{T,max}}$.

We simultaneously fit the three sets of data with their corresponding model functions $\mathcalboondox{A}(\theta_{{\rm{cav}},x})$, $\eta_{\rm{dip}}(\theta_{{\rm{cav}},x})$, $\epsilon_{\HT}(\theta_{{\rm{cav}},x})$ by minimizing the (unweighted) sum of all least-square residuals (see Fig.~\ref{fig:opencavity}a and Supplemental Material~\cite{Note3}). In addition to the independently measured constants $R_\HT = 150\pm15\,\mu$m, $R_\ULT$, $L_{\rm{cav}} = 16.9\pm0.5\,\mu$m, $d_{f}= 2.0\pm0.4\,\mu$m, $w_f = 2.60\pm0.05$, $\mathcal{T_{\rm{HT,850\,\rm{nm}}}}$ and $\mathcal{L_{\rm{sum,850\,\rm{nm}}}}$, the model functions depend on four experimentally unknown or inaccurately constrained constants, which we use as free parameters of the fit. These are the alignment angle $\theta_{{\rm{cav}},y} \approx 0$ along the y-axis, the absolute zero position of the scan of the alignment angle along the x-axis, the orientation $\phi$ of the fiber mirror decentration axis relative to the x-axis ($d_{f,x}= d_f \cdot \cos(\phi)$, $d_{f,y} = d_f\cdot \sin(\phi))$ and the absolute calibration constant $C_T$ of the mode matching efficiency. 
We find good agreement between the experimental data and the theoretical model for the hybrid cavity (see Fig.~\ref{fig:opencavity}). Significantly better agreement is found for a fiber mirror decentration of $d_f= 2.7\,\mu$m (see Supplemental Material~\cite{Note3}), which could indicate that our characterization of the fiber mirror by interferometric imaging ($d_{f}= 2.0\pm0.4\,\mu$m) underestimates the fiber-mirror decentration. 

Our investigation makes it possible to determine and therefore optimize individual mode matching efficiencies $\epsilon_{\HT}$, $\epsilon_{\LT}$ using the easily accessible quantities $\mathcalboondox{A}$ and  $\eta_{\rm{dip}}$ from cavity reflection measurements. 
\section{Conclusion}
\label{sec:conclusion}
The results presented in this article address some important practical questions that arise during the experimental implementation of high finesse fiber Fabry-Perot cavities: How can optimal fiber cavity alignment be achieved and how can the individual mode matching efficiencies be characterized? How should optical fiber cavities be constructed and stabilized to fulfill their potential for miniaturization and integration into robust scientific and technical devices?

The first two questions we address with a fiber mode matching calculation that relates the alignment-dependent mode-matching efficiency to the easily measurable dip in the reflected light power at the cavity resonance. 

The latter question we explore by investigating two different fiber resonators architectures, a traditional piezo-mechanically actuated resonator and a novel, intrinsically rigid fiber cavity.  

The piezo-actuated fiber cavity provides maximum tunability of the resonance frequency but entails experimental complexity. % Temperature stability and thermal effects are in general not an issue for the actively stabilized cavity as drifts can effectively compensated by piezo-feedback.
The rigid fiber cavity, on the other hand, exhibits a high degree of passive stability but relies on thermal control for frequency tuning and drift compensation. Our analysis of the thermal-response of FFPCs and the demonstration of long-term stabilization, both by active external feedback by means of a heating laser and by self-locking, show that precise locking of rigid high finesse FFPCs cavities can be realized.
We believe that, due to their simplicity, compactness and robustness, rigid fiber cavities could be a promising solution for miniaturized, integratable and scalable CQED devices and further applications of FFPC with similar requirements.  \\
Note: After the preparation of this manuscript, we have been made aware of closely related results on the mode matching of fiber Fabry-Perot cavities~\cite{Bick2016}.

The work has been supported by the Bundesministerium f\"ur Bildung und Forschung (BMBF, Verbund Q.com-Q) and by funds of the Integrated Project SIQS of the European Commission. J.G., S.K.A., M.M. and L.R. acknowledge support by the European Commission training network CCQED, the DAAD, the Bonn-Cologne Graduate School of Physics and Astronomy and the Alexander von Humboldt Foundation, respectively.


\begin{thebibliography}{36}%
\makeatletter
\providecommand \@ifxundefined [1]{%
 \@ifx{#1\undefined}
}%
\providecommand \@ifnum [1]{%
 \ifnum #1\expandafter \@firstoftwo
 \else \expandafter \@secondoftwo
 \fi
}%
\providecommand \@ifx [1]{%
 \ifx #1\expandafter \@firstoftwo
 \else \expandafter \@secondoftwo
 \fi
}%
\providecommand \natexlab [1]{#1}%
\providecommand \enquote  [1]{``#1''}%
\providecommand \bibnamefont  [1]{#1}%
\providecommand \bibfnamefont [1]{#1}%
\providecommand \citenamefont [1]{#1}%
\providecommand \href@noop [0]{\@secondoftwo}%
\providecommand \href [0]{\begingroup \@sanitize@url \@href}%
\providecommand \@href[1]{\@@startlink{#1}\@@href}%
\providecommand \@@href[1]{\endgroup#1\@@endlink}%
\providecommand \@sanitize@url [0]{\catcode `\\12\catcode `\$12\catcode
  `\&12\catcode `\#12\catcode `\^12\catcode `\_12\catcode `\%12\relax}%
\providecommand \@@startlink[1]{}%
\providecommand \@@endlink[0]{}%
\providecommand \url  [0]{\begingroup\@sanitize@url \@url }%
\providecommand \@url [1]{\endgroup\@href {#1}{\urlprefix }}%
\providecommand \urlprefix  [0]{URL }%
\providecommand \Eprint [0]{\href }%
\providecommand \doibase [0]{http://dx.doi.org/}%
\providecommand \selectlanguage [0]{\@gobble}%
\providecommand \bibinfo  [0]{\@secondoftwo}%
\providecommand \bibfield  [0]{\@secondoftwo}%
\providecommand \translation [1]{[#1]}%
\providecommand \BibitemOpen [0]{}%
\providecommand \bibitemStop [0]{}%
\providecommand \bibitemNoStop [0]{.\EOS\space}%
\providecommand \EOS [0]{\spacefactor3000\relax}%
\providecommand \BibitemShut  [1]{\csname bibitem#1\endcsname}%
\let\auto@bib@innerbib\@empty
%</preamble>
\bibitem [{\citenamefont {Kuhn}\ \emph {et~al.}(2002)\citenamefont {Kuhn},
  \citenamefont {Hennrich},\ and\ \citenamefont {Rempe}}]{Kuhn2002}%
  \BibitemOpen
  \bibfield  {author} {\bibinfo {author} {\bibfnamefont {A.}~\bibnamefont
  {Kuhn}}, \bibinfo {author} {\bibfnamefont {M.}~\bibnamefont {Hennrich}}, \
  and\ \bibinfo {author} {\bibfnamefont {G.}~\bibnamefont {Rempe}},\ }\href
  {\doibase 10.1103/PhysRevLett.89.067901} {\bibfield  {journal} {\bibinfo
  {journal} {Phys. Rev. Lett.}\ }\textbf {\bibinfo {volume} {89}},\ \bibinfo
  {pages} {067901} (\bibinfo {year} {2002})}\BibitemShut {NoStop}%
\bibitem [{\citenamefont {Ritter}\ \emph {et~al.}(2012)\citenamefont {Ritter},
  \citenamefont {N\"{o}lleke}, \citenamefont {Hahn}, \citenamefont {Reiserer},
  \citenamefont {Neuzner}, \citenamefont {Uphoff}, \citenamefont {M\"{u}cke},
  \citenamefont {Figueroa}, \citenamefont {Bochmann},\ and\ \citenamefont
  {Rempe}}]{Ritter2012}%
  \BibitemOpen
  \bibfield  {author} {\bibinfo {author} {\bibfnamefont {S.}~\bibnamefont
  {Ritter}}, \bibinfo {author} {\bibfnamefont {C.}~\bibnamefont {N\"{o}lleke}},
  \bibinfo {author} {\bibfnamefont {C.}~\bibnamefont {Hahn}}, \bibinfo {author}
  {\bibfnamefont {A.}~\bibnamefont {Reiserer}}, \bibinfo {author}
  {\bibfnamefont {A.}~\bibnamefont {Neuzner}}, \bibinfo {author} {\bibfnamefont
  {M.}~\bibnamefont {Uphoff}}, \bibinfo {author} {\bibfnamefont
  {M.}~\bibnamefont {M\"{u}cke}}, \bibinfo {author} {\bibfnamefont
  {E.}~\bibnamefont {Figueroa}}, \bibinfo {author} {\bibfnamefont
  {J.}~\bibnamefont {Bochmann}}, \ and\ \bibinfo {author} {\bibfnamefont
  {G.}~\bibnamefont {Rempe}},\ }\href {\doibase 10.1038/nature11023} {\bibfield
   {journal} {\bibinfo  {journal} {Nature}\ }\textbf {\bibinfo {volume}
  {484}},\ \bibinfo {pages} {195} (\bibinfo {year} {2012})}\BibitemShut
  {NoStop}%
\bibitem [{\citenamefont {Purdy}\ \emph {et~al.}(2010)\citenamefont {Purdy},
  \citenamefont {Brooks}, \citenamefont {Botter}, \citenamefont {Brahms},
  \citenamefont {Ma},\ and\ \citenamefont {Stamper-Kurn}}]{Purdy2010}%
  \BibitemOpen
  \bibfield  {author} {\bibinfo {author} {\bibfnamefont {T.~P.}\ \bibnamefont
  {Purdy}}, \bibinfo {author} {\bibfnamefont {D.~W.~C.}\ \bibnamefont
  {Brooks}}, \bibinfo {author} {\bibfnamefont {T.}~\bibnamefont {Botter}},
  \bibinfo {author} {\bibfnamefont {N.}~\bibnamefont {Brahms}}, \bibinfo
  {author} {\bibfnamefont {Z.-Y.}\ \bibnamefont {Ma}}, \ and\ \bibinfo {author}
  {\bibfnamefont {D.~M.}\ \bibnamefont {Stamper-Kurn}},\ }\href
  {http://dx.doi.org/10.1103/PhysRevLett.105.133602} {\bibfield  {journal}
  {\bibinfo  {journal} {Phys. Rev. Lett.}\ }\textbf {\bibinfo {volume} {105}},\
  \bibinfo {pages} {133602} (\bibinfo {year} {2010})}\BibitemShut {NoStop}%
\bibitem [{\citenamefont {Colombe}\ \emph {et~al.}(2007)\citenamefont
  {Colombe}, \citenamefont {Steinmetz}, \citenamefont {Dubois}, \citenamefont
  {Linke}, \citenamefont {Hunger},\ and\ \citenamefont
  {Reichel}}]{Colombe2007}%
  \BibitemOpen
  \bibfield  {author} {\bibinfo {author} {\bibfnamefont {Y.}~\bibnamefont
  {Colombe}}, \bibinfo {author} {\bibfnamefont {T.}~\bibnamefont {Steinmetz}},
  \bibinfo {author} {\bibfnamefont {G.}~\bibnamefont {Dubois}}, \bibinfo
  {author} {\bibfnamefont {F.}~\bibnamefont {Linke}}, \bibinfo {author}
  {\bibfnamefont {D.}~\bibnamefont {Hunger}}, \ and\ \bibinfo {author}
  {\bibfnamefont {J.}~\bibnamefont {Reichel}},\ }\href
  {http://dx.doi.org/10.1038/nature06331} {\bibfield  {journal} {\bibinfo
  {journal} {Nature}\ }\textbf {\bibinfo {volume} {450}},\ \bibinfo {pages}
  {272} (\bibinfo {year} {2007})}\BibitemShut {NoStop}%
\bibitem [{\citenamefont {Muller}\ \emph {et~al.}(2010)\citenamefont {Muller},
  \citenamefont {Flagg}, \citenamefont {Lawall},\ and\ \citenamefont
  {Solomon}}]{Muller2010}%
  \BibitemOpen
  \bibfield  {author} {\bibinfo {author} {\bibfnamefont {A.}~\bibnamefont
  {Muller}}, \bibinfo {author} {\bibfnamefont {E.~B.}\ \bibnamefont {Flagg}},
  \bibinfo {author} {\bibfnamefont {J.~R.}\ \bibnamefont {Lawall}}, \ and\
  \bibinfo {author} {\bibfnamefont {G.~S.}\ \bibnamefont {Solomon}},\ }\href
  {\doibase 10.1364/OL.35.002293} {\bibfield  {journal} {\bibinfo  {journal}
  {Opt. Lett.}\ }\textbf {\bibinfo {volume} {35}},\ \bibinfo {pages} {2293}
  (\bibinfo {year} {2010})}\BibitemShut {NoStop}%
\bibitem [{\citenamefont {Hunger}\ \emph {et~al.}(2012)\citenamefont {Hunger},
  \citenamefont {Deutsch}, \citenamefont {Barbour}, \citenamefont {Warburton},\
  and\ \citenamefont {Reichel}}]{Hunger2012}%
  \BibitemOpen
  \bibfield  {author} {\bibinfo {author} {\bibfnamefont {D.}~\bibnamefont
  {Hunger}}, \bibinfo {author} {\bibfnamefont {C.}~\bibnamefont {Deutsch}},
  \bibinfo {author} {\bibfnamefont {R.~J.}\ \bibnamefont {Barbour}}, \bibinfo
  {author} {\bibfnamefont {R.~J.}\ \bibnamefont {Warburton}}, \ and\ \bibinfo
  {author} {\bibfnamefont {J.}~\bibnamefont {Reichel}},\ }\href
  {http://dx.doi.org/10.1063/1.3679721} {\bibfield  {journal} {\bibinfo
  {journal} {AIP Advances}\ }\textbf {\bibinfo {volume} {2}},\ \bibinfo {eid}
  {012119} (\bibinfo {year} {2012})}\BibitemShut {NoStop}%
\bibitem [{\citenamefont {Hunger}\ \emph {et~al.}(2010)\citenamefont {Hunger},
  \citenamefont {Steinmetz}, \citenamefont {Colombe}, \citenamefont {Deutsch},
  \citenamefont {H{\"a}nsch},\ and\ \citenamefont {Reichel}}]{Hunger2010}%
  \BibitemOpen
  \bibfield  {author} {\bibinfo {author} {\bibfnamefont {D.}~\bibnamefont
  {Hunger}}, \bibinfo {author} {\bibfnamefont {T.}~\bibnamefont {Steinmetz}},
  \bibinfo {author} {\bibfnamefont {Y.}~\bibnamefont {Colombe}}, \bibinfo
  {author} {\bibfnamefont {C.}~\bibnamefont {Deutsch}}, \bibinfo {author}
  {\bibfnamefont {T.~W.}\ \bibnamefont {H{\"a}nsch}}, \ and\ \bibinfo {author}
  {\bibfnamefont {J.}~\bibnamefont {Reichel}},\ }\href
  {http://dx.doi.org/10.1088/1367-2630/12/6/065038} {\bibfield  {journal}
  {\bibinfo  {journal} {New Journal of Physics}\ }\textbf {\bibinfo {volume}
  {12}},\ \bibinfo {pages} {065038} (\bibinfo {year} {2010})}\BibitemShut
  {NoStop}%
\bibitem [{\citenamefont {Brandst\"{a}tter}\ \emph {et~al.}(2013)\citenamefont
  {Brandst\"{a}tter}, \citenamefont {McClung}, \citenamefont {Sch\"{u}ppert},
  \citenamefont {Casabone}, \citenamefont {Friebe}, \citenamefont {Stute},
  \citenamefont {Schmidt}, \citenamefont {Deutsch}, \citenamefont {Reichel},
  \citenamefont {Blatt},\ and\ \citenamefont {Northup}}]{Brandstatter2013}%
  \BibitemOpen
  \bibfield  {author} {\bibinfo {author} {\bibfnamefont {B.}~\bibnamefont
  {Brandst\"{a}tter}}, \bibinfo {author} {\bibfnamefont {A.}~\bibnamefont
  {McClung}}, \bibinfo {author} {\bibfnamefont {K.}~\bibnamefont
  {Sch\"{u}ppert}}, \bibinfo {author} {\bibfnamefont {B.}~\bibnamefont
  {Casabone}}, \bibinfo {author} {\bibfnamefont {K.}~\bibnamefont {Friebe}},
  \bibinfo {author} {\bibfnamefont {A.}~\bibnamefont {Stute}}, \bibinfo
  {author} {\bibfnamefont {P.~O.}\ \bibnamefont {Schmidt}}, \bibinfo {author}
  {\bibfnamefont {C.}~\bibnamefont {Deutsch}}, \bibinfo {author} {\bibfnamefont
  {J.}~\bibnamefont {Reichel}}, \bibinfo {author} {\bibfnamefont
  {R.}~\bibnamefont {Blatt}}, \ and\ \bibinfo {author} {\bibfnamefont {T.~E.}\
  \bibnamefont {Northup}},\ }\href {\doibase 10.1063/1.4838696} {\bibfield
  {journal} {\bibinfo  {journal} {Rev. of Sci. Instr.}\ }\textbf {\bibinfo
  {volume} {84}},\ \bibinfo {pages} {123104} (\bibinfo {year}
  {2013})}\BibitemShut {NoStop}%
\bibitem [{\citenamefont {Takahashi}\ \emph {et~al.}(2014)\citenamefont
  {Takahashi}, \citenamefont {Morphew}, \citenamefont {Oru\v{c}evi\'{c}},
  \citenamefont {Noguchi}, \citenamefont {Kassa},\ and\ \citenamefont
  {Keller}}]{Takahashi2014}%
  \BibitemOpen
  \bibfield  {author} {\bibinfo {author} {\bibfnamefont {H.}~\bibnamefont
  {Takahashi}}, \bibinfo {author} {\bibfnamefont {J.}~\bibnamefont {Morphew}},
  \bibinfo {author} {\bibfnamefont {F.}~\bibnamefont {Oru\v{c}evi\'{c}}},
  \bibinfo {author} {\bibfnamefont {A.}~\bibnamefont {Noguchi}}, \bibinfo
  {author} {\bibfnamefont {E.}~\bibnamefont {Kassa}}, \ and\ \bibinfo {author}
  {\bibfnamefont {M.}~\bibnamefont {Keller}},\ }\href
  {http://dx.doi.org/10.1364/OE.22.031317} {\bibfield  {journal} {\bibinfo
  {journal} {Opt. Express}\ }\textbf {\bibinfo {volume} {22}},\ \bibinfo
  {pages} {31317} (\bibinfo {year} {2014})}\BibitemShut {NoStop}%
\bibitem [{\citenamefont {Uphoff}\ \emph {et~al.}(2015)\citenamefont {Uphoff},
  \citenamefont {Brekenfeld}, \citenamefont {Rempe},\ and\ \citenamefont
  {Ritter}}]{Uphoff2015}%
  \BibitemOpen
  \bibfield  {author} {\bibinfo {author} {\bibfnamefont {M.}~\bibnamefont
  {Uphoff}}, \bibinfo {author} {\bibfnamefont {M.}~\bibnamefont {Brekenfeld}},
  \bibinfo {author} {\bibfnamefont {G.}~\bibnamefont {Rempe}}, \ and\ \bibinfo
  {author} {\bibfnamefont {S.}~\bibnamefont {Ritter}},\ }\href
  {http://dx.doi.org/10.1088/1367-2630/17/1/013053} {\bibfield  {journal}
  {\bibinfo  {journal} {New Journal of Physics}\ }\textbf {\bibinfo {volume}
  {17}},\ \bibinfo {pages} {013053} (\bibinfo {year} {2015})}\BibitemShut
  {NoStop}%
\bibitem [{\citenamefont {Steiner}\ \emph {et~al.}(2013)\citenamefont
  {Steiner}, \citenamefont {Meyer}, \citenamefont {Deutsch}, \citenamefont
  {Reichel},\ and\ \citenamefont {K\"ohl}}]{Steiner2013}%
  \BibitemOpen
  \bibfield  {author} {\bibinfo {author} {\bibfnamefont {M.}~\bibnamefont
  {Steiner}}, \bibinfo {author} {\bibfnamefont {H.~M.}\ \bibnamefont {Meyer}},
  \bibinfo {author} {\bibfnamefont {C.}~\bibnamefont {Deutsch}}, \bibinfo
  {author} {\bibfnamefont {J.}~\bibnamefont {Reichel}}, \ and\ \bibinfo
  {author} {\bibfnamefont {M.}~\bibnamefont {K\"ohl}},\ }\href
  {http://dx.doi.org/10.1103/PhysRevLett.110.043003} {\bibfield  {journal}
  {\bibinfo  {journal} {Phys. Rev. Lett.}\ }\textbf {\bibinfo {volume} {110}},\
  \bibinfo {pages} {043003} (\bibinfo {year} {2013})}\BibitemShut {NoStop}%
\bibitem [{\citenamefont {Toninelli}\ \emph {et~al.}(2010)\citenamefont
  {Toninelli}, \citenamefont {Delley}, \citenamefont {St\"{o}ferle},
  \citenamefont {Renn}, \citenamefont {G\"{o}tzinger},\ and\ \citenamefont
  {Sandoghdar}}]{Toninelli2010}%
  \BibitemOpen
  \bibfield  {author} {\bibinfo {author} {\bibfnamefont {C.}~\bibnamefont
  {Toninelli}}, \bibinfo {author} {\bibfnamefont {Y.}~\bibnamefont {Delley}},
  \bibinfo {author} {\bibfnamefont {T.}~\bibnamefont {St\"{o}ferle}}, \bibinfo
  {author} {\bibfnamefont {A.}~\bibnamefont {Renn}}, \bibinfo {author}
  {\bibfnamefont {S.}~\bibnamefont {G\"{o}tzinger}}, \ and\ \bibinfo {author}
  {\bibfnamefont {V.}~\bibnamefont {Sandoghdar}},\ }\href
  {http://dx.doi.org/10.1063/1.3456559} {\bibfield  {journal} {\bibinfo
  {journal} {Applied Physics Letters}\ }\textbf {\bibinfo {volume} {97}},\
  \bibinfo {pages} {021107} (\bibinfo {year} {2010})}\BibitemShut {NoStop}%
\bibitem [{\citenamefont {Muller}\ \emph {et~al.}(2009)\citenamefont {Muller},
  \citenamefont {Flagg}, \citenamefont {Metcalfe}, \citenamefont {Lawall},\
  and\ \citenamefont {Solomon}}]{Muller2009}%
  \BibitemOpen
  \bibfield  {author} {\bibinfo {author} {\bibfnamefont {A.}~\bibnamefont
  {Muller}}, \bibinfo {author} {\bibfnamefont {E.~B.}\ \bibnamefont {Flagg}},
  \bibinfo {author} {\bibfnamefont {M.}~\bibnamefont {Metcalfe}}, \bibinfo
  {author} {\bibfnamefont {J.}~\bibnamefont {Lawall}}, \ and\ \bibinfo {author}
  {\bibfnamefont {G.~S.}\ \bibnamefont {Solomon}},\ }\href
  {http://dx.doi.org/10.1063/1.3245311} {\bibfield  {journal} {\bibinfo
  {journal} {Applied Physics Letters}\ }\textbf {\bibinfo {volume} {95}},\
  \bibinfo {eid} {173101} (\bibinfo {year} {2009})}\BibitemShut {NoStop}%
\bibitem [{\citenamefont {Albrecht}\ \emph {et~al.}(2013)\citenamefont
  {Albrecht}, \citenamefont {Bommer}, \citenamefont {Deutsch}, \citenamefont
  {Reichel},\ and\ \citenamefont {Becher}}]{Albrecht2013}%
  \BibitemOpen
  \bibfield  {author} {\bibinfo {author} {\bibfnamefont {R.}~\bibnamefont
  {Albrecht}}, \bibinfo {author} {\bibfnamefont {A.}~\bibnamefont {Bommer}},
  \bibinfo {author} {\bibfnamefont {C.}~\bibnamefont {Deutsch}}, \bibinfo
  {author} {\bibfnamefont {J.}~\bibnamefont {Reichel}}, \ and\ \bibinfo
  {author} {\bibfnamefont {C.}~\bibnamefont {Becher}},\ }\href {\doibase
  10.1103/PhysRevLett.110.243602} {\bibfield  {journal} {\bibinfo  {journal}
  {Phys. Rev. Lett.}\ }\textbf {\bibinfo {volume} {110}},\ \bibinfo {pages}
  {243602} (\bibinfo {year} {2013})}\BibitemShut {NoStop}%
\bibitem [{\citenamefont {Flowers-Jacobs}\ \emph {et~al.}(2012)\citenamefont
  {Flowers-Jacobs}, \citenamefont {Hoch}, \citenamefont {Sankey}, \citenamefont
  {Kashkanova}, \citenamefont {Jayich}, \citenamefont {Deutsch}, \citenamefont
  {Reichel},\ and\ \citenamefont {Harris}}]{Flowers2012}%
  \BibitemOpen
  \bibfield  {author} {\bibinfo {author} {\bibfnamefont {N.~E.}\ \bibnamefont
  {Flowers-Jacobs}}, \bibinfo {author} {\bibfnamefont {S.~W.}\ \bibnamefont
  {Hoch}}, \bibinfo {author} {\bibfnamefont {J.~C.}\ \bibnamefont {Sankey}},
  \bibinfo {author} {\bibfnamefont {A.}~\bibnamefont {Kashkanova}}, \bibinfo
  {author} {\bibfnamefont {A.~M.}\ \bibnamefont {Jayich}}, \bibinfo {author}
  {\bibfnamefont {C.}~\bibnamefont {Deutsch}}, \bibinfo {author} {\bibfnamefont
  {J.}~\bibnamefont {Reichel}}, \ and\ \bibinfo {author} {\bibfnamefont
  {J.~G.~E.}\ \bibnamefont {Harris}},\ }\href
  {http://dx.doi.org/10.1063/1.4768779} {\bibfield  {journal} {\bibinfo
  {journal} {Applied Physics Letters}\ }\textbf {\bibinfo {volume} {101}},\
  \bibinfo {eid} {221109} (\bibinfo {year} {2012})}\BibitemShut {NoStop}%
\bibitem [{Note1()}]{Note1}%
  \BibitemOpen
  \bibinfo {note} {IVG\protect \textsuperscript {\textregistered }
  Cu800.}\BibitemShut {Stop}%
\bibitem [{Note2()}]{Note2}%
  \BibitemOpen
  \bibinfo {note} {EPO-TEK\protect \textsuperscript {\textregistered }
  OG116-31}\BibitemShut {NoStop}%
\bibitem [{\citenamefont {Mabuchi}\ \emph {et~al.}(1999)\citenamefont
  {Mabuchi}, \citenamefont {Ye},\ and\ \citenamefont {Kimble}}]{Mabuchi1999}%
  \BibitemOpen
  \bibfield  {author} {\bibinfo {author} {\bibfnamefont {H.}~\bibnamefont
  {Mabuchi}}, \bibinfo {author} {\bibfnamefont {J.}~\bibnamefont {Ye}}, \ and\
  \bibinfo {author} {\bibfnamefont {H.~J.}\ \bibnamefont {Kimble}},\ }\href
  {http://dx.doi.org/10.1007/s003400050751} {\bibfield  {journal} {\bibinfo
  {journal} {Applied Physics B}\ }\textbf {\bibinfo {volume} {68}},\ \bibinfo
  {pages} {1095} (\bibinfo {year} {1999})}\BibitemShut {NoStop}%
\bibitem [{\citenamefont {Whittaker}\ \emph {et~al.}(1985)\citenamefont
  {Whittaker}, \citenamefont {Gehrtz},\ and\ \citenamefont
  {Bjorklund}}]{Whittaker1985}%
  \BibitemOpen
  \bibfield  {author} {\bibinfo {author} {\bibfnamefont {E.~A.}\ \bibnamefont
  {Whittaker}}, \bibinfo {author} {\bibfnamefont {M.}~\bibnamefont {Gehrtz}}, \
  and\ \bibinfo {author} {\bibfnamefont {G.~C.}\ \bibnamefont {Bjorklund}},\
  }\href {\doibase 10.1364/JOSAB.2.001320} {\bibfield  {journal} {\bibinfo
  {journal} {J. Opt. Soc. Am. B}\ }\textbf {\bibinfo {volume} {2}},\ \bibinfo
  {pages} {1320} (\bibinfo {year} {1985})}\BibitemShut {NoStop}%
\bibitem [{\citenamefont {Wieman}\ and\ \citenamefont
  {Gilbert}(1982)}]{Wieman1982}%
  \BibitemOpen
  \bibfield  {author} {\bibinfo {author} {\bibfnamefont {C.~E.}\ \bibnamefont
  {Wieman}}\ and\ \bibinfo {author} {\bibfnamefont {S.~L.}\ \bibnamefont
  {Gilbert}},\ }\href {\doibase 10.1364/OL.7.000480} {\bibfield  {journal}
  {\bibinfo  {journal} {Opt. Lett.}\ }\textbf {\bibinfo {volume} {7}},\
  \bibinfo {pages} {480} (\bibinfo {year} {1982})}\BibitemShut {NoStop}%
\bibitem [{\citenamefont {Shaddock}\ \emph {et~al.}(1999)\citenamefont
  {Shaddock}, \citenamefont {Gray},\ and\ \citenamefont
  {McClelland}}]{Shaddock1999}%
  \BibitemOpen
  \bibfield  {author} {\bibinfo {author} {\bibfnamefont {D.~A.}\ \bibnamefont
  {Shaddock}}, \bibinfo {author} {\bibfnamefont {M.~B.}\ \bibnamefont {Gray}},
  \ and\ \bibinfo {author} {\bibfnamefont {D.~E.}\ \bibnamefont {McClelland}},\
  }\href {\doibase 10.1364/OL.24.001499} {\bibfield  {journal} {\bibinfo
  {journal} {Opt. Lett.}\ }\textbf {\bibinfo {volume} {24}},\ \bibinfo {pages}
  {1499} (\bibinfo {year} {1999})}\BibitemShut {NoStop}%
\bibitem [{Note3()}]{Note3}%
  \BibitemOpen
  \bibinfo {note} {The Supplemental Material is included in the source code of
  the article that can be downloaded from its arXiv page.}\BibitemShut {Stop}%
\bibitem [{\citenamefont {Reimann}\ \emph {et~al.}(2015)\citenamefont
  {Reimann}, \citenamefont {Alt}, \citenamefont {Kampschulte}, \citenamefont
  {Macha}, \citenamefont {Ratschbacher}, \citenamefont {Thau}, \citenamefont
  {Yoon},\ and\ \citenamefont {Meschede}}]{Reimann2015}%
  \BibitemOpen
  \bibfield  {author} {\bibinfo {author} {\bibfnamefont {R.}~\bibnamefont
  {Reimann}}, \bibinfo {author} {\bibfnamefont {W.}~\bibnamefont {Alt}},
  \bibinfo {author} {\bibfnamefont {T.}~\bibnamefont {Kampschulte}}, \bibinfo
  {author} {\bibfnamefont {T.}~\bibnamefont {Macha}}, \bibinfo {author}
  {\bibfnamefont {L.}~\bibnamefont {Ratschbacher}}, \bibinfo {author}
  {\bibfnamefont {N.}~\bibnamefont {Thau}}, \bibinfo {author} {\bibfnamefont
  {S.}~\bibnamefont {Yoon}}, \ and\ \bibinfo {author} {\bibfnamefont
  {D.}~\bibnamefont {Meschede}},\ }\href
  {http://dx.doi.org/10.1103/PhysRevLett.114.023601} {\bibfield  {journal}
  {\bibinfo  {journal} {Phys. Rev. Lett.}\ }\textbf {\bibinfo {volume} {114}},\
  \bibinfo {pages} {023601} (\bibinfo {year} {2015})}\BibitemShut {NoStop}%
\bibitem [{\citenamefont {Wardle}(1999)}]{Wardle1999}%
  \BibitemOpen
  \bibfield  {author} {\bibinfo {author} {\bibfnamefont {D.}~\bibnamefont
  {Wardle}},\ }\emph {\bibinfo {title} {Raman scattering in optical fibres}},\
  \href {http://hdl.handle.net/2292/433} {Ph.D. thesis},\ \bibinfo  {school}
  {University of Auckland} (\bibinfo {year} {1999})\BibitemShut {NoStop}%
\bibitem [{\citenamefont {Martin}\ and\ \citenamefont {Ye}(2012)}]{Martin2012}%
  \BibitemOpen
  \bibfield  {author} {\bibinfo {author} {\bibfnamefont {M.~J.}\ \bibnamefont
  {Martin}}\ and\ \bibinfo {author} {\bibfnamefont {J.}~\bibnamefont {Ye}},\
  }\href@noop {} {\emph {\bibinfo {title} {High-precision Laser Stabilization
  via Optical Cavities}}},\ Optical Coatings and Thermal Noise in Precision
  Measurement\ (\bibinfo  {publisher} {Cambridge University Press},\ \bibinfo
  {year} {2012})\BibitemShut {NoStop}%
\bibitem [{Note4()}]{Note4}%
  \BibitemOpen
  \bibinfo {note} {TRA-BOND\protect \textsuperscript {\textregistered }
  F112}\BibitemShut {NoStop}%
\bibitem [{\citenamefont {Carmon}\ \emph {et~al.}(2004)\citenamefont {Carmon},
  \citenamefont {Yang},\ and\ \citenamefont {Vahala}}]{Carmon2004}%
  \BibitemOpen
  \bibfield  {author} {\bibinfo {author} {\bibfnamefont {T.}~\bibnamefont
  {Carmon}}, \bibinfo {author} {\bibfnamefont {L.}~\bibnamefont {Yang}}, \ and\
  \bibinfo {author} {\bibfnamefont {K.}~\bibnamefont {Vahala}},\ }\href
  {\doibase 10.1364/OPEX.12.004742} {\bibfield  {journal} {\bibinfo  {journal}
  {Opt. Express}\ }\textbf {\bibinfo {volume} {12}},\ \bibinfo {pages} {4742}
  (\bibinfo {year} {2004})}\BibitemShut {NoStop}%
\bibitem [{\citenamefont {An}\ \emph {et~al.}(1997)\citenamefont {An},
  \citenamefont {Sones}, \citenamefont {Fang-Yen}, \citenamefont {Dasari},\
  and\ \citenamefont {Feld}}]{An1997}%
  \BibitemOpen
  \bibfield  {author} {\bibinfo {author} {\bibfnamefont {K.}~\bibnamefont
  {An}}, \bibinfo {author} {\bibfnamefont {B.~A.}\ \bibnamefont {Sones}},
  \bibinfo {author} {\bibfnamefont {C.}~\bibnamefont {Fang-Yen}}, \bibinfo
  {author} {\bibfnamefont {R.~R.}\ \bibnamefont {Dasari}}, \ and\ \bibinfo
  {author} {\bibfnamefont {M.~S.}\ \bibnamefont {Feld}},\ }\href
  {http://dx.doi.org/10.1364/OL.22.001433} {\bibfield  {journal} {\bibinfo
  {journal} {Opt. Lett.}\ }\textbf {\bibinfo {volume} {22}},\ \bibinfo {pages}
  {1433} (\bibinfo {year} {1997})}\BibitemShut {NoStop}%
\bibitem [{\citenamefont {Asenbaum}\ \emph {et~al.}(2013)\citenamefont
  {Asenbaum}, \citenamefont {Kuhn}, \citenamefont {Nimmrichter}, \citenamefont
  {Sezer},\ and\ \citenamefont {Arndt}}]{Asenbaum2013}%
  \BibitemOpen
  \bibfield  {author} {\bibinfo {author} {\bibfnamefont {P.}~\bibnamefont
  {Asenbaum}}, \bibinfo {author} {\bibfnamefont {S.}~\bibnamefont {Kuhn}},
  \bibinfo {author} {\bibfnamefont {S.}~\bibnamefont {Nimmrichter}}, \bibinfo
  {author} {\bibfnamefont {U.}~\bibnamefont {Sezer}}, \ and\ \bibinfo {author}
  {\bibfnamefont {M.}~\bibnamefont {Arndt}},\ }\href
  {http://dx.doi.org/10.1038/ncomms3743} {\bibfield  {journal} {\bibinfo
  {journal} {Nat. Commun.}\ }\textbf {\bibinfo {volume} {4}},\ \bibinfo
  {pages} {2743} (\bibinfo {year}
  {2013})}\BibitemShut {NoStop}%
\bibitem [{\citenamefont {Crooks}\ \emph {et~al.}(2006)\citenamefont {Crooks},
  \citenamefont {Cagnoli}, \citenamefont {Fejer}, \citenamefont {Harry},
  \citenamefont {Hough}, \citenamefont {Khuri-Yakub}, \citenamefont {Penn},
  \citenamefont {Route}, \citenamefont {Rowan}, \citenamefont {Sneddon},
  \citenamefont {Wygant},\ and\ \citenamefont {Yaralioglu}}]{Crooks2006}%
  \BibitemOpen
  \bibfield  {author} {\bibinfo {author} {\bibfnamefont {D.~R.~M.}\
  \bibnamefont {Crooks}}, \bibinfo {author} {\bibfnamefont {G.}~\bibnamefont
  {Cagnoli}}, \bibinfo {author} {\bibfnamefont {M.~M.}\ \bibnamefont {Fejer}},
  \bibinfo {author} {\bibfnamefont {G.}~\bibnamefont {Harry}}, \bibinfo
  {author} {\bibfnamefont {J.}~\bibnamefont {Hough}}, \bibinfo {author}
  {\bibfnamefont {B.~T.}\ \bibnamefont {Khuri-Yakub}}, \bibinfo {author}
  {\bibfnamefont {S.}~\bibnamefont {Penn}}, \bibinfo {author} {\bibfnamefont
  {R.}~\bibnamefont {Route}}, \bibinfo {author} {\bibfnamefont
  {S.}~\bibnamefont {Rowan}}, \bibinfo {author} {\bibfnamefont {P.~H.}\
  \bibnamefont {Sneddon}}, \bibinfo {author} {\bibfnamefont {I.~O.}\
  \bibnamefont {Wygant}}, \ and\ \bibinfo {author} {\bibfnamefont {G.~G.}\
  \bibnamefont {Yaralioglu}},\ }\href
  {http://dx.doi.org/10.1088/0264-9381/23/15/014} {\bibfield  {journal}
  {\bibinfo  {journal} {Classical and Quantum Gravity}\ }\textbf {\bibinfo
  {volume} {23}},\ \bibinfo {pages} {4953} (\bibinfo {year}
  {2006})}\BibitemShut {NoStop}%
\bibitem [{Note5()}]{Note5}%
  \BibitemOpen
  \bibinfo {note} {Wolfram Mathematica\protect \textsuperscript
  {\textregistered }}\BibitemShut {NoStop}%
\bibitem [{\citenamefont {Joyce}\ and\ \citenamefont
  {DeLoach}(1984)}]{Joyce1984}%
  \BibitemOpen
  \bibfield  {author} {\bibinfo {author} {\bibfnamefont {W.~B.}\ \bibnamefont
  {Joyce}}\ and\ \bibinfo {author} {\bibfnamefont {B.~C.}\ \bibnamefont
  {DeLoach}},\ }\href {\doibase 10.1364/AO.23.004187} {\bibfield  {journal}
  {\bibinfo  {journal} {Appl. Opt.}\ }\textbf {\bibinfo {volume} {23}},\
  \bibinfo {pages} {4187} (\bibinfo {year} {1984})}\BibitemShut {NoStop}%
\bibitem [{Note6()}]{Note6}%
  \BibitemOpen
  \bibinfo {note} {This is evident for the overlap integral of Gaussian beams,
  where for the complex spatial mode functions the relation $\psi _{i,\protect
  \mathaccentV {vec}17E{k}}(x,y,z)= \psi _{i,-\protect \mathaccentV
  {vec}17E{k}}^*(x,y,z)$ holds and hence $\DOTSI \intop \ilimits@ _{-\infty
  }^{\infty }\DOTSI \intop \ilimits@ _{-\infty }^{\infty }\psi ^*_{1,\protect
  \mathaccentV {vec}17E{k}_1}(x,y,z)\psi _{2,\protect \mathaccentV
  {vec}17E{k}_2}(x,y,z)\protect \tmspace +\thinmuskip {.1667em} dxdy = (\DOTSI
  \intop \ilimits@ _{-\infty }^{\infty }\DOTSI \intop \ilimits@ _{-\infty
  }^{\infty }\psi ^*_{1,-\protect \mathaccentV {vec}17E{k}_1}(x,y,z)\psi
  _{2,-\protect \mathaccentV {vec}17E{k}_2}(x,y,z)\protect \tmspace
  +\thinmuskip {.1667em} dxdy)^*$ (see Supplemental Material~\cite{Note3} for
  details).}\BibitemShut {Stop}%
\bibitem [{\citenamefont {Hood}\ \emph {et~al.}(2001)\citenamefont {Hood},
  \citenamefont {Kimble},\ and\ \citenamefont {Ye}}]{Hood2001}%
  \BibitemOpen
  \bibfield  {author} {\bibinfo {author} {\bibfnamefont {C.~J.}\ \bibnamefont
  {Hood}}, \bibinfo {author} {\bibfnamefont {H.~J.}\ \bibnamefont {Kimble}}, \
  and\ \bibinfo {author} {\bibfnamefont {J.}~\bibnamefont {Ye}},\ }\href
  {\doibase 10.1103/PhysRevA.64.033804} {\bibfield  {journal} {\bibinfo
  {journal} {Phys. Rev. A}\ }\textbf {\bibinfo {volume} {64}},\ \bibinfo
  {pages} {033804} (\bibinfo {year} {2001})}\BibitemShut {NoStop}%
\bibitem [{\citenamefont {Siegman}(1986)}]{Siegman1986}%
  \BibitemOpen
  \bibfield  {author} {\bibinfo {author} {\bibfnamefont {A.~E.}\ \bibnamefont
  {Siegman}},\ }\href@noop {} {\emph {\bibinfo {title} {Lasers}}}\ (\bibinfo
  {publisher} {University Science Books},\ \bibinfo {year} {1986})\BibitemShut
  {NoStop}%
\bibitem [{\citenamefont {Bick}\ \emph {et~al.}(2016)\citenamefont {Bick},
  \citenamefont {Staarmann}, \citenamefont {Christoph}, \citenamefont
  {Hellmig}, \citenamefont {Heinze}, \citenamefont {Sengstock},\ and\
  \citenamefont {Becker}}]{Bick2016}%
  \BibitemOpen
  \bibfield  {author} {\bibinfo {author} {\bibfnamefont {A.}~\bibnamefont
  {Bick}}, \bibinfo {author} {\bibfnamefont {C.}~\bibnamefont {Staarmann}},
  \bibinfo {author} {\bibfnamefont {P.}~\bibnamefont {Christoph}}, \bibinfo
  {author} {\bibfnamefont {O.}~\bibnamefont {Hellmig}}, \bibinfo {author}
  {\bibfnamefont {J.}~\bibnamefont {Heinze}}, \bibinfo {author} {\bibfnamefont
  {K.}~\bibnamefont {Sengstock}}, \ and\ \bibinfo {author} {\bibfnamefont
  {C.}~\bibnamefont {Becker}},\ }\href {"http://dx.doi.org/10.1063/1.4939046"}
  {\bibfield  {journal} {\bibinfo  {journal} {Review of Scientific
  Instruments}\ }\textbf {\bibinfo {volume} {87}},\ \bibinfo {eid} {013102}
  (\bibinfo {year} {2016})}\BibitemShut {NoStop}%
\end{thebibliography}
\end{document}